\documentclass[prd,preprintnumbers,nofootinbib,superscriptaddress,onecolumn,notitlepage]
{revtex4}

\usepackage{cancel}
\usepackage{amsmath,mathrsfs,amscd,graphicx}
\usepackage{mciteplus}
\usepackage{multirow}
\usepackage{color}
\usepackage{cancel}
\usepackage{rotating}
\usepackage{slashed}
\usepackage{subfig}
\usepackage{textcomp}
\usepackage[final]{pdfpages}
\usepackage{array}
\usepackage{float}
\usepackage{url}
\usepackage{textcomp}
\usepackage{amsfonts, latexsym, epsfig}
\usepackage{bm}
\usepackage{times}
\usepackage{epsfig}
\usepackage{amssymb}
\usepackage{tikz}
\usepackage{slashed}
\usepackage{hyperref}
\usepackage{verbatim}
\usepackage[normalem]{ulem}
\usepackage[utf8]{inputenc}
\usepackage{diagbox}
\usepackage{titlesec}
\def\beq{\begin{equation}}
\def\eeq{\end{equation}}
\def\be{\begin{equation}}
\def\ee{\end{equation}}
\def\bea{\begin{eqnarray}}
\def\eea{\end{eqnarray}}

\definecolor{dpmagenta}{rgb}{0.8, 0.0, 0.8}

\allowdisplaybreaks[3]
\begin{document}
\title{Electroweak corrections to Higgs boson production via Z Z fusion at the future LHeC}
\author{Hanying Xiong~}
\email{21736003@zju.edu.cn}
\affiliation{School of Physics, Hangzhou Normal University, Hangzhou, Zhejiang 311121, China}
\author{Hongsheng Hou~}
\email{hshou@hznu.edu.cn}
\affiliation{School of Physics, Hangzhou Normal University, Hangzhou, Zhejiang 311121, China}
\author{Zhuoni Qian~}
\email{sdaly@126.com}
\affiliation{School of Physics, Hangzhou Normal University, Hangzhou, Zhejiang 311121, China}
\author{Qingjun Xu~}
\email{xuqingjun@hznu.edu.cn}
\affiliation{School of Physics, Hangzhou Normal University, Hangzhou, Zhejiang 311121, China}
\author{Bowen Wang~}
\email{bowenw@hznu.edu.cn}
\affiliation{School of Physics, Hangzhou Normal University, Hangzhou, Zhejiang 311121, China}

\begin{abstract}
	An important mechanism for production of the Higgs boson at the prospective Large Hadron-electron Collider (LHeC) is via neutral current (NC)
	weak boson fusion (WBF) processes. Aside from its role in measurements of Higgs couplings within the standard model, 
	this production mode is particularly useful in searchings of Higgs decays into invisble particles in various models for the
	Higg portal dark matter. In this work we compute the electroweak corrections for the NC WBF at the LHeC up to the 1-loop level. 
	For a center-of-mass energy of 1.98 TeV, the magnitudes of the relative corrections for the total cross section at next-to-leading (NLO) order are 
	respectively 8\% and 17\%, in the two renormalization schemes we use. The NLO terms also distort various distributions (notably, those for
	Higgs and electron observables) computed at the leading order. Along with our previous treatment of the charge current processes, this paper
	completes the calulation of the NLO EW effects for the dominant Higgs production modes at the LHeC.
	
\end{abstract}

\maketitle

\section{Introduction}
Various Higgs boson production mechanisims have been explored to study the properties of the particle, especially over the decade
after its discovery~\cite{Aad:2012tfa,Chatrchyan:2012xdj}. 
Among these, the production of the Higgs via weak boson fusion (WBF) has unique kinematic 
characteristics that have been succesfully utilized in the determination of Higgs couplings with other particles. 
For instance, the measurement of Higgs-gauge boson couplings was found to be sensitive to the WBF production mode 
(as well as to other modes such as the gluon-gluon fusion, or ggH)
at the LHC~\cite{ATLAS:2020qdt,CMS:2020gsy}. The combined analysis reaches an accuracy at the percent level.
In addition, this channel also plays an important role in probing the $H\rightarrow \tau^+\tau^-$ decay~\cite{CMS:2017zyp,ATLAS:2018ynr}

To avoid large QCD background, the measurement of Higgs Yukawa couplings with light quarks is proposed to be performed at 
lepton-hadron colliders such as the future LHeC~\cite{LHeCStudyGroup:2012zhm}\footnote{Of course, there are many other motivations for building
the LHeC. For more details see e.g., Ref.~\cite{LHeCStudyGroup:2012zhm}}, 
which is planned to run at a center-of-mass energy of $1\sim 2$ TeV 
with the current proton beam at the LHC scattering on an additional beam of electrons. Studies have already shown the potential of probing
the $H$-$b$ Yukawa coupling at the LHeC~\cite{Han:2009pe,LHeCStudyGroup:2012zhm,LHeC:2020van}, 
where Higgs bosons are predominantly produced via WBF. Positive results are also obtained in restricting the 
$H$-$c$ Yukawa and triple-higgs-self couplings~\cite{Li:2019xwd,Li:2019jba,LHeC:2020van}. These advancements
give rise to the need for including higher order corrections of the cross sections in the simulations of the relevant processes. 
In fact, the QCD~\cite{Jager:2010zm} and part of the electro-weak~\cite{Blumlein:1992eh} corrections to the Higgs WBF production on 
e-p colliders at 1-loop level were computed even before many of these phenomenological studies. 

In a previous paper~\cite{Wang:2022awk}, 
we\footnote{Two of the three authors of this reference are in the author list of the current paper.} treated the Higgs WBF production via
W boson fusion with the full EW corrections at 1 loop (next-to-leading order, or NLO).
At the tree level, the charge current (CC) processes account for approximately 80\% of the total cross section 
for Higgs production at the energy regime of
the LHeC. However, the neutral current (NC) processes via Z Z fusion is not negligible, and moreover, has its own merit because the final state
electrons can be observed in about half of the scattering events (in contrast to the missing energy carried by neutrinos in CC events).
The NC final state will typically contain two visible fermions with large rapidity difference and invariant mass, 
as well as the products from the fusion subprocesses near
the central region of the rapidity gap. This configuration not only provides a prospect of restricting the $H$-$b$ coupling 
with the NC WBF~\cite{Han:2009pe}, but is also useful particularly when probing final states with invisible fusion products, 
thanks to the kinematic handles provided by the forward and backward fermions\footnote{The classification into ``forward'', ``backward'', 
and ``central'' may be changed by boosts along the beams, but it is possible to distinguish various final states 
with observables such as the rapidities of the particles.}.

In fact, studies at the LHC are the first to make use of the WBF mode in dark matter searches~\cite{Bai:2011wz,Ghosh:2012ep,Dutta:2012xe,Delannoy:2013ata,Bernaciak:2014pna}, where the forward and backward jets can be
tagged in both CC and NC processes. In particular, Higgs decays to dark matter particles allowed 
by various extentions~\cite{Arkani-Hamed:1998wuz,Eboli:2000ze,Djouadi:2011aa,Englert:2011yb,Mambrini:2011ik,Baek:2012se,Belanger:2013kya,Curtin:2013fra} of the standard model (SM) are investigated experimentally 
by ATLAS~\cite{ATLAS:2019cid} and CMS~\cite{CMS:2016dhk} groups. 
In these searches, the ggH mode with jets is not preferable since it suffers from large background, e.g., $Z\rightarrow \nu\overline{\nu}$ 
+ jets in the central region. Due to lack of discriminative power from the event topology, strict restrictions on the jets have to be imposed 
in order to suppress the background, which substantially lowers the signal rate. 
The search in VH (V denotes W or Z) production, where V decays leptonically, gives better result, 
but WBF turns out to be the most sensitive channel in identifying invisible decays of the Higgs, 
and sets the lowest upper limit (18\% by CMS~\cite{CMS:2022qva,CMS:2018yfx} and 28\% by ATLAS~\cite{ATLAS:2015gvj,ATLAS:2021pdg}, 
at 95\% confidence level) for the branching ratio of these
decays. 

On the other hand, it is shown that more stringent limit is to be expected at the LHeC,
where the background involves Z and W bosons produced via purely electro-weak interactions~\cite{Tang:2015uha}.  
In contrast, the background for the LHC analysis
includes  Z/W +jets, with jets produced by strong interactions, which is less clean and controllable.
However, the W W fusion processes cannot be used in dark matter searches at the LHeC as the single jet signal is not quite differentiable from 
the charge current deep inelastic scattering.
The study in Ref.~\cite{Tang:2015uha} focuses then on the NC processes and reaches an upper limit of 6\% at $2\sigma$ level.

Encouraged by these results, we are to carry out in this paper the caculation of the EW corrections for the NC WBF processes at the LHeC. 
We shall follow closely the method and notations of our previous treatment of the CC processes~\cite{Wang:2022awk}, and organize the rest
of this paper as follows. Section~\ref{higgs_xsecs} gives the details of the calculation. Section~\ref{result} shows numerically the significance
of the NLO corrections for the NC processes. In Section~\ref{conclusion} we discuss the result and conclude the paper.

\section{Calculation of the processes}
\label{higgs_xsecs}
In describing the treatment of the NC processes, we shall keep to the minimum the discussion that 
repeats the CC calculation~\cite{Wang:2022awk}, and focus more on the new features of the NC processes.
 \subsection{Leading order (LO) contribution}
\label{LO}
\begin{figure}[h]
\centering
    \includegraphics[scale=0.5]{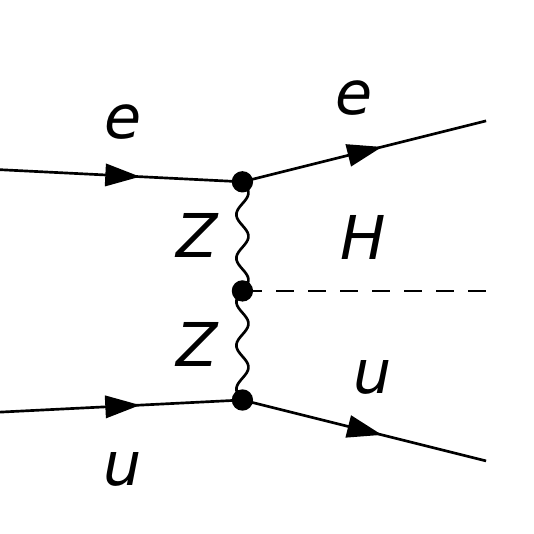} \qquad
	\caption{A representative LO graph for NC WBF at the LHeC.}
\label{born}
\end{figure}

The LO cross section for the NC processes is computed according to
\begin{equation}
\begin{aligned}
	\sigma^{LO}=\sum_{q}\int_0^1 d\eta_q f_q(\eta_q,\mu_F) \frac{1}{2\hat{s}}d\Phi_3 B(P_A,p_q),    
\label{eq:LOxsec}
\end{aligned}
\end{equation}
where the squared amplitude $B(P_A,p_q)$ for the born process $e q \rightarrow e q H$, along with the corresponding flux factor, 
is convolved with the proper parton distribution function. The result is then summed over all quark (antiquark) flavors, 
except for $b$, $\bar{b}$, $t$, and $\bar{t}$, whose contributions are marginal. A LO graph with an
incoming u quark is shown in Fig.~\ref{born}.

\subsection{NLO EW corrections}
\label{NLO}
The calculation of NLO EW corrections involves cancellation of various types of singularities and must be treated with care. 
As with the CC case, here we still work in the framework of the dipole subtraction method~\cite{Catani:1996vz,Schonherr:2017qcj}, 
and organize the calculation into numerical integrations over 4- and 3-particle final states respectively.
 
\subsubsection{4-particle final states}
\label{4-part}
\begin{figure}[H]
\centering
    \includegraphics[scale=0.7]{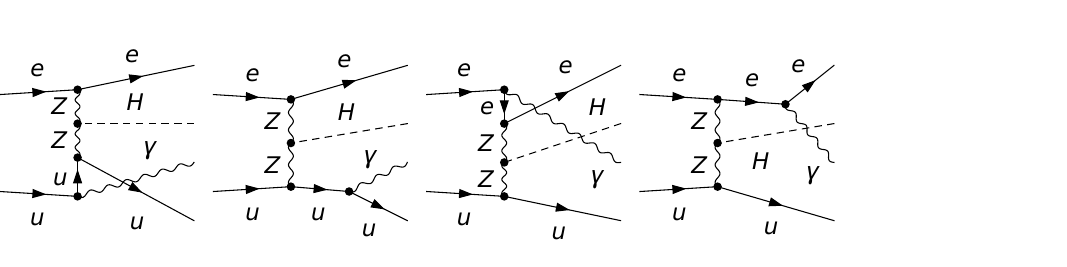} \qquad
    \includegraphics[scale=0.7]{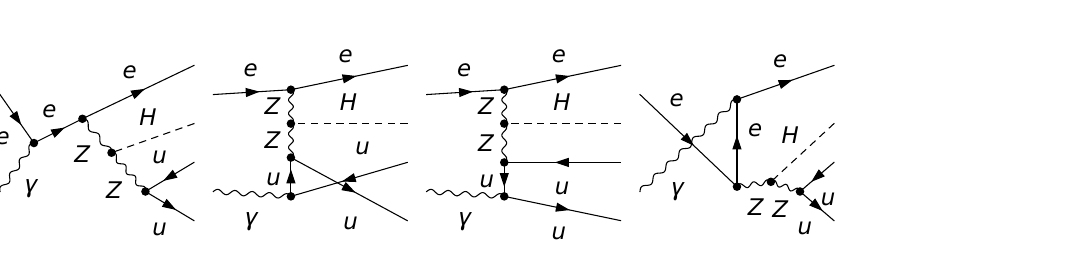}\qquad
\caption{Representative real-emission graphs for NC WBF at the LHeC. The first and second lines 
	correspond respectively to the processes $e^-+q \rightarrow e^-+\gamma+q+H$ 
	and $e^-+\gamma \rightarrow e^-+q+\bar{q}+H$, in which $q=u$.}
\label{real}
\end{figure}

Neglecting the contribution from $t$, $b$, and their antiparticles gives 48 NLO real emission graphs 
with a photon in the initial or final state, some of which are shown in Fig.~\ref{real}. These tree graphs, as well as the LO graph, 
are evaluated with the packages {\sc FeynArts}, {\sc FormCalc}, and {\sc LoopTools}~\cite{Hahn:2000kx,Hahn:1998yk}. 
The photon induced graphs may involve 
resonant propagators and are handled with the complex-mass-scheme~\cite{Denner:1999gp,Denner:2005fg,Denner:2006ic} in the calculation.

The subtraction term of the real emission processes have the form
\begin{equation}
\begin{aligned}
|\mathcal{M}_{sub}|^2=\sum_{i, j}\sum_k\mathcal{D}_{ij,k}+\sum_{i, j}\sum_a\mathcal{D}_{ij}^a+\sum_{a,j}\sum_{k\neq j}\mathcal{D}_{j,k}^a+\sum_{a,j}\sum_{b\neq a}\mathcal{D}_{j}^{a,b}.
\label{eq:realdip}
\end{aligned}
\end{equation}
where the labeling of the emitter, emittee and spectator follows the convention in Refs.~\cite{Schonherr:2017qcj,Wang:2022awk}. 
Summing over all contributing flavor combinations gives complete sets of dipoles
for the quark (photon) induced processes, whose detailed forms are collected in Appendix~\ref{app:4-particle-final-states}. 
Note that the first term in Eq.~\ref{eq:realdip} represents dipoles with both the splitting particles and the spectator from the final state. 
This type of dipoles is not present in the CC processes.

The contribution from 4-particle final states is then obtained by subtracting the singularities from the real emission cross section:
\begin{equation}
\begin{aligned}
\sigma_4^{NLO}=\sum_b \int_0^1d\eta_bf_b(\eta_b,\mu_F)\frac{1}{2\hat{s}}d\Phi_4\left\{|\mathcal{M}_R^b|^2F^{(2)}(p_1,p_2,p_3,p_4;P_A,p_b)-|\mathcal{M}_{sub}^b|^2\right\},
\label{eq:realsub}
\end{aligned}
\end{equation}
with a sum over the parton flavor $b$ that collides with the incoming electron. 
A particular jet algorithm is implemented via the function $F^{(2)}$ as well as an implicit $F^{(1)}$ multiplied with
the subtraction term in order to ensure IR and collinear safety for jet observables.\footnote{One
could also consider collinear unsafe observables~\cite{Dittmaier:2008md}. Distributions in this case are enhanced 
logarithmically by final state radiations in collinear regions. 
For simplicity, however, this kind of observables is not explored in this study.}

\subsubsection{3-particle final states}

\begin{figure}[h]
\centering
    \includegraphics[scale=0.7]{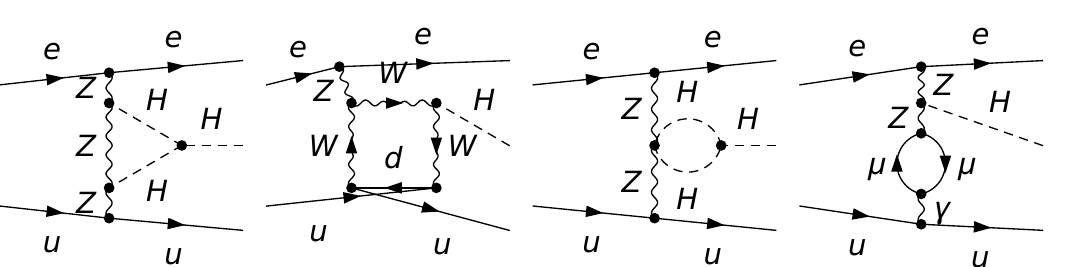}\qquad
    \includegraphics[scale=0.7]{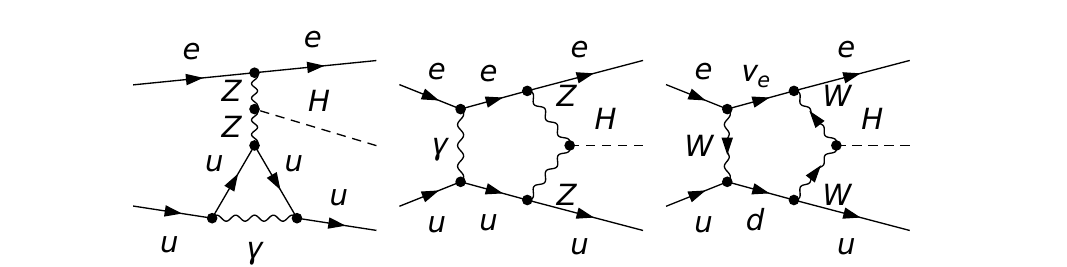}\qquad
\caption{Representative 1-loop graphs for NC WBF at the LHeC.}
\label{loop}
\end{figure}

The contribution of the 3-particle final states comes from loop-graph terms combined with integrated dipoles as well as 
collinear counter terms in a certain factorization scheme. After a reorganization of terms, one arrives at an expression
\begin{equation}
\begin{aligned}
	\sigma_3^{NLO}=&\sum_{b}\int  d\eta_b f_b(\eta_b,\mu_F)\bigg\{\int \frac{1}{2\hat{s}}d\Phi_3^{(4)}\Big[V_{ab}(\Phi_3,P_A,p_b)+B_{ab}(\Phi_3,P_A,p_b) \boldsymbol{I}^b(\epsilon,\mu^2)\Big]_{\epsilon =0}\\
&+\sum_{a'}\int dx_a\int\frac{1}{2\hat{s}}d\Phi_3^{(4)} B_{a'b}(\Phi_3^{(4)},x_a P_A,p_b) \Big[\boldsymbol{K}^b_{aa'}(x_a)+\boldsymbol{P}^b_{aa'}(x_a;\mu_F^2)\Big]\\
	&+\sum_{b'}\int dx_b\int\frac{1}{2\hat{s}}d\Phi_3^{(4)} B_{ab'}(\Phi_3^{(4)},P_A,x_b p_b) \Big[\boldsymbol{K}^a_{bb'}(x_b)+\boldsymbol{P}^a_{bb'}(x_b;\mu_F^2)\Big]\bigg\}\\
	&+\sum_{b}\int d\eta_a d\eta_b f^{\mathcal{O}(\alpha)}_e(\eta_a,\mu_F)f_b(\eta_b,\mu_F)\int \frac{1}{2\hat{s}}d\Phi_3^{(4)}B_{ab}(\Phi_3^{(4)},p_a,p_b).
\label{eq:virtualsub}
\end{aligned}
\end{equation}
The terms with $V_{ab}$ is from 1-loop graphs. 
In this calculation, 2280 1-loop graphs (see examples in Fig.~\ref{loop}) are produced and evaluated
by the program {\sc MadGraph5\_aMC@NLO}~\cite{Alwall:2014hca}. Renormalization of parameters is done in 
both $G_{\mu}$ and $\alpha(M_Z)$ schemes~\cite{Denner:1991kt,Denner:2019vbn}.
The explicit form of the $\boldsymbol{I}$,$\boldsymbol{K}$ and $\boldsymbol{P}$~\cite{Catani:1996vz,Catani:2002hc,Schonherr:2017qcj} terms
are given in Appendix~\ref{app:3-particle-final-states}. The last line of Eq.~\ref{eq:virtualsub} is obtained by factorization of 
an electron distribution function to simplify the treatment of the initial state radiations, where the $\mathcal{O}(\alpha)$ electron distribution
reads~\cite{Liu:2021jfp}
\begin{equation}
\begin{aligned}
f^{\mathcal{O}(\alpha)}_e(x,\mu_F)=\frac{\alpha}{2\pi}\Big[\frac{1+x^2}{1-x}\ln\frac{\mu_F^2}{(1-x)^2m_e^2}\Big]_+.
\label{eq:electron_pdf}
\end{aligned}
\end{equation}
The factorization of both the electron and partons from the proton is carried out in the $\overline{\mbox{MS}}$ scheme.
A jet definition, as remarked at the end of section~\ref{4-part}, is applied to the 3-particle final states 
by an implicit function $F^{(1)}$ both at LO and NLO.

\section{Numerical result}
{\label{result}}
The beam energies and SM parameters for this calculation are taken as
\begin{equation}
\begin{aligned}
	&E_e=140 \; \mbox{GeV}, \qquad E_p=7 \; \mbox{TeV},\\
	&G_{\mu}=1.16639\times 10^{-5} \; \mbox{GeV}^{-2}, \qquad \alpha_{G_{\mu}}=1/132.5,\qquad \alpha(M_Z)=1/128.93,\\
	&M_W=80.419 \; \mbox{GeV}, \qquad \Gamma_W=2.09291 \; \mbox{GeV},\qquad M_Z=91.188 \; \mbox{GeV}, \\
	& \Gamma_Z=2.49877 \; \mbox{GeV},\qquad c_W^2=1-s_W^2=\frac{M_W^2}{M_Z^2},\\
    &m_e=0.510998928 \; \mbox{MeV}, \qquad M_H=125 \; \mbox{GeV}.
\end{aligned}
\end{equation}
A unit CKM matrix is used for simplicity, and we take $M_W$ for the renormalization and factorization scales. All quark and electron masses are
set to zero in the scattering amplitudes, which reduces substantially the number of graphs to be evaluated. 
The 3- and 4- particle phase spaces are generated in our own program that 
relies on the Vegas Monte-Carlo algorithm from the {\sc Cuba} library~\cite{Hahn:2004fe} for numerical integration. 
Checks similar to those in the CC case are done until a consistent and stable result is obtained.


Now we show our numerical results of cross sections computed with the setup above.
The total integrated cross sections listed in Table~\ref{int-xsec} are about 20\% of those in CC case. 
However, the corrections from the NLO terms relative to the LO ones are respectively -8\% and -17\% in $G_{\mu}$ and $\alpha(M_Z)$ schemes,
which are very close to the results for CC processes.
Furthermore, the  decomposition of the total cross sections into
various contributions at LO and NLO in two schemes shows a pattern of convergence very similar to the CC result,
which reflects once again the renormalization scheme independence of the physical result. At NLO, the ratio between the contributions
from 3- and 4- particle final states increases by a factor of 4 as compared with the CC case. 
This is rouphly proportional to the increase of the ratio between the number of loop- and real-graphs in NC WBF.

\begin{table}[h]
\begin{center}
\begin{tabular}{|c|c|c|c|c|}
\hline
	Schemes & total & LO & 3-particle & 4-particle \\
\hline
	$G_{\mu}$ &  45.02 & 48.78 &  -3.85 & 0.09 \\
\hline
	$\alpha(M_Z)$ & 43.81 & 52.95 & -9.24 & 0.10 \\
\hline
\end{tabular}
\end{center}
	\caption{Integrated cross sections in fb for NC WBF at the LHeC at LO and NLO, computed in two renormalization schemes $G_{\mu}$ and $\alpha(M_Z)$. The electron and proton beam energies are $140$~GeV and $7$~TeV, respectively.}
\label{int-xsec}
\end{table}

For differential distributions, we first construct quark jets and electron observables 
according to the $k_T$ algorithm~\cite{Catani:1992zp,Ellis:1993tq,Blazey:2000qt} with the parameter $D=0.8$.
Hence the electron observables we shall discuss below receive contribution from the photon momentum in the soft and collinear limits.
Next, we apply selection cuts to the jet with the largest 
transverse momentum (tagging jet), as well as to the electron observables, requiring
\begin{equation}
\begin{aligned}
\label{eq:cuts}
	&p_{T}^{j}>30~\text{GeV}, \qquad  p_{T}^{e}>25~\text{GeV}, \\
	  & \qquad -5<\eta_{j}, \, \eta_{e}<-5,
\end{aligned}
\end{equation}
where these quantities are defined with respect to the forward direction in which the incoming electron is moving. 
Cuts on transverse momenta are inherited from the CC calculation to maintain a typical event shape of WBF. The requirement on
rapidities is loosened as compared with the CC case, because the NC signal is much smaller
at the LHeC and could not afford a substantial loss. For the same reason we no longer restrict 
 the invant mass of the Higgs-tagging jet (or of the Higgs-electron) system~\cite{Han:2009pe}.

\begin{figure}[h!]
\centering
\subfloat[$p_T^{j}$ distribution]{\includegraphics[scale=0.4]{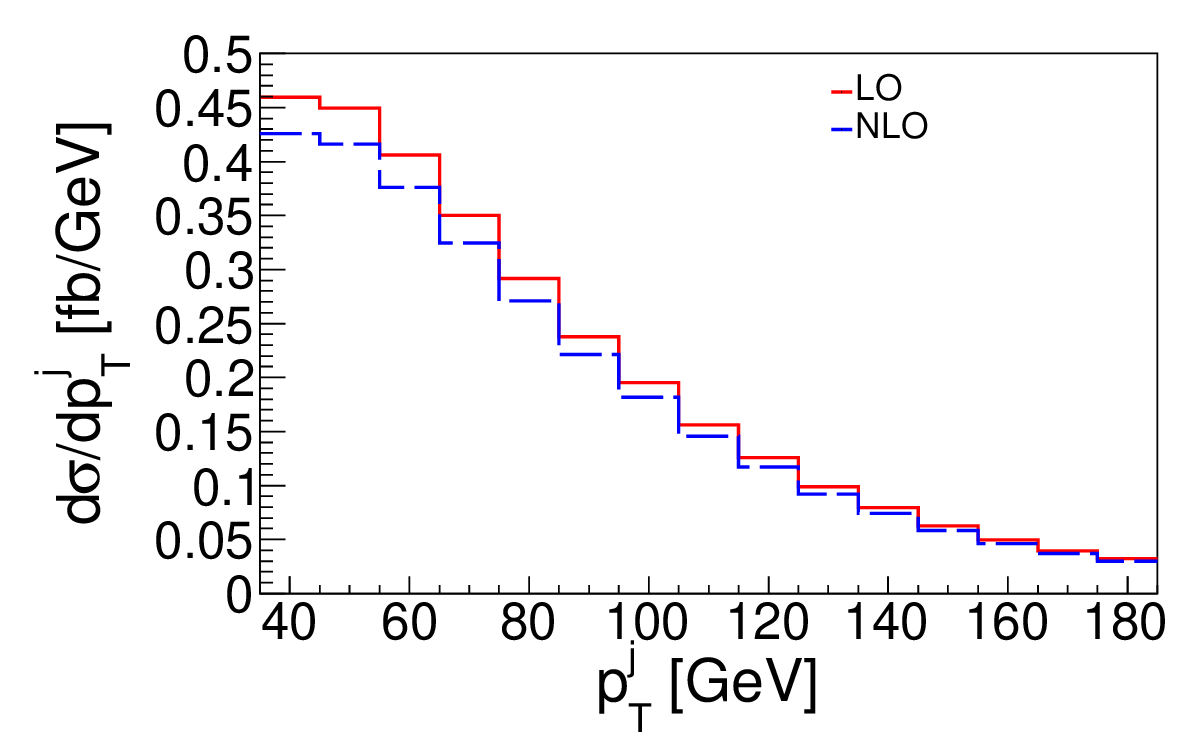}}\qquad
\subfloat[$K$ factor]{\includegraphics[scale=0.4]{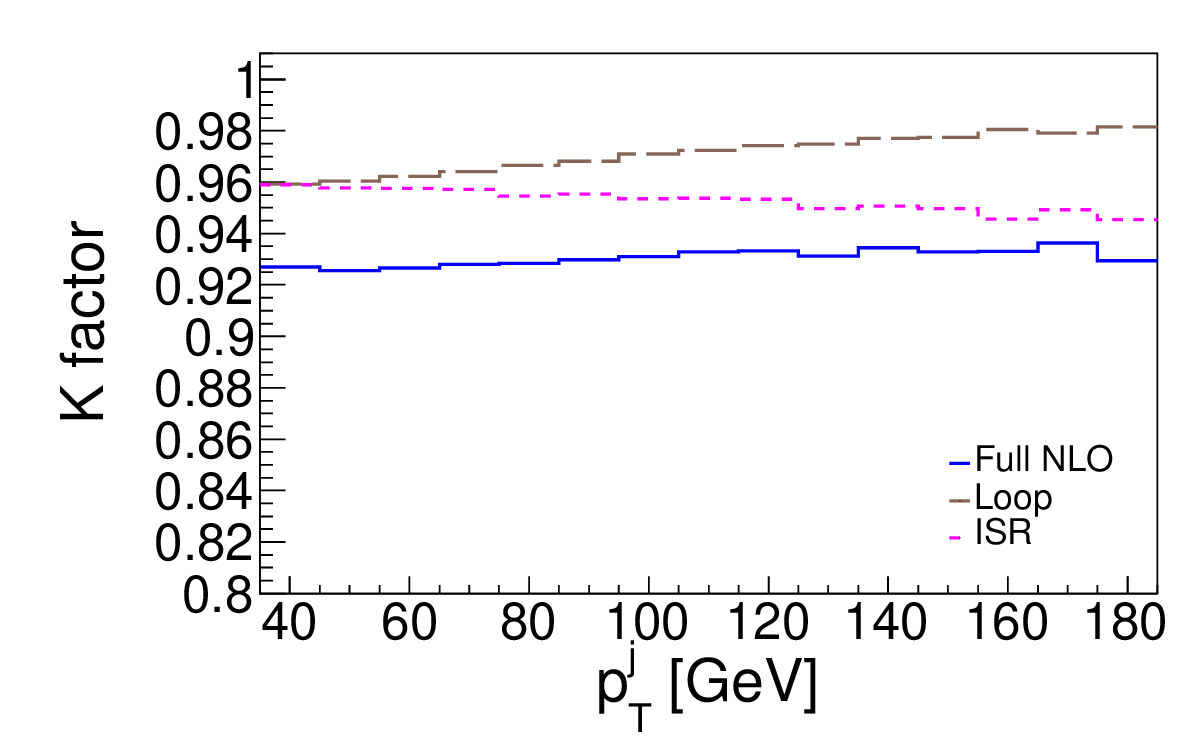}}\qquad \\
	\caption{Distribution in the transverse momentum $p_T^j$ of the tagging jet (a), and the corresponding $K$ factors (b).}
\label{ptj}
\end{figure}

To present the result, we compare various cross sections computed to LO and NLO (LO + NLO corrections).
We shall work in the $G_{\mu}$ scheme as it leads to a faster convergence of the perturbative calculation.
Fig.~\ref{ptj} shows the transverse momentum distribution and the corresponding K factors of the tagging jet. 
The curves on the left drop rapidly at large $p_T^j$, and the
NLO correction is negative throughout the entire spectrum shown. 
The $p_T^j$ of a typical event is located around $\mathcal{O}(M_Z)$, as is expected for the WBF.
The K factor plot on the right shows the distribution computed up to NLO divided by the corresponding one at LO. 
Also displayed are The K factors with solely the loop contribution and with solely the initial state radiations (ISR) off the electron (i.e., the 
contribution from the $\mathcal{O}(\alpha)$ electron distribution function), respectively, at NLO. Corrections from both sources are about
-5\% and change slowly with increasing $p_T^j$. However, their sum gives almost a constant K factor $\sim 0.93$, as shown by the blue curve.

\begin{figure}[h!]
\centering
\subfloat[$\eta_{j}$ distribution]{\includegraphics[scale=0.4]{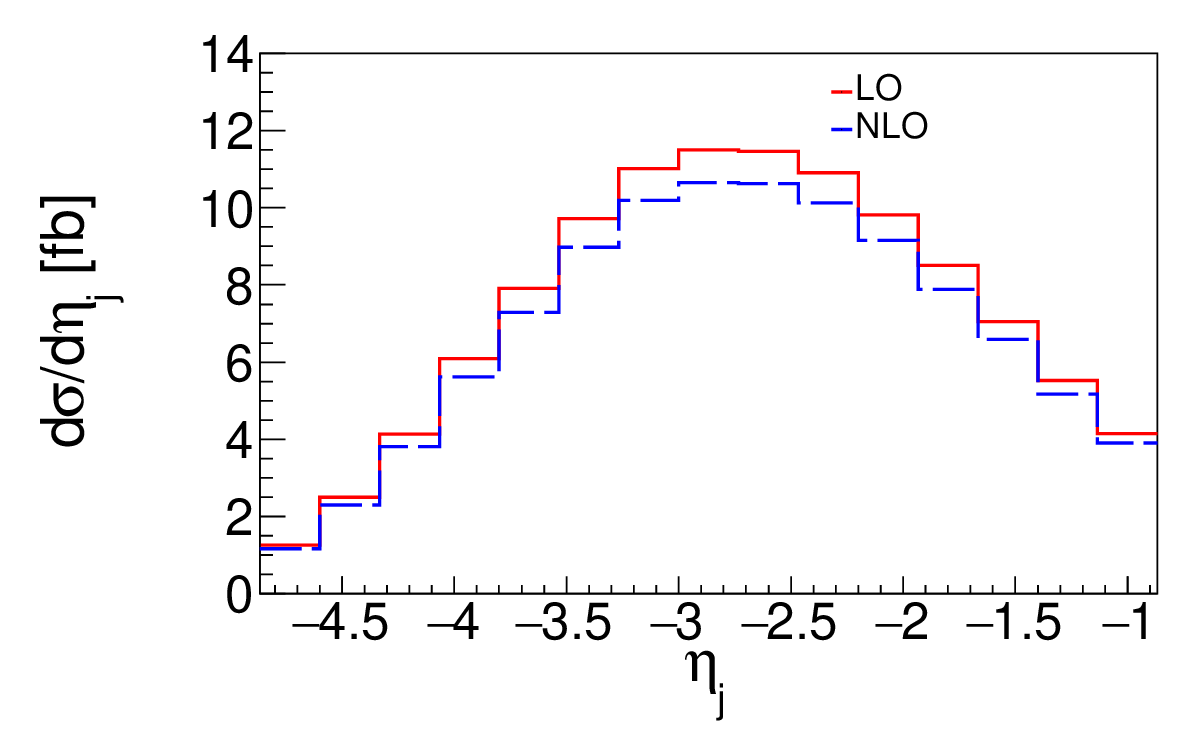}}\qquad
\subfloat[$K$ factor]{\includegraphics[scale=0.4]{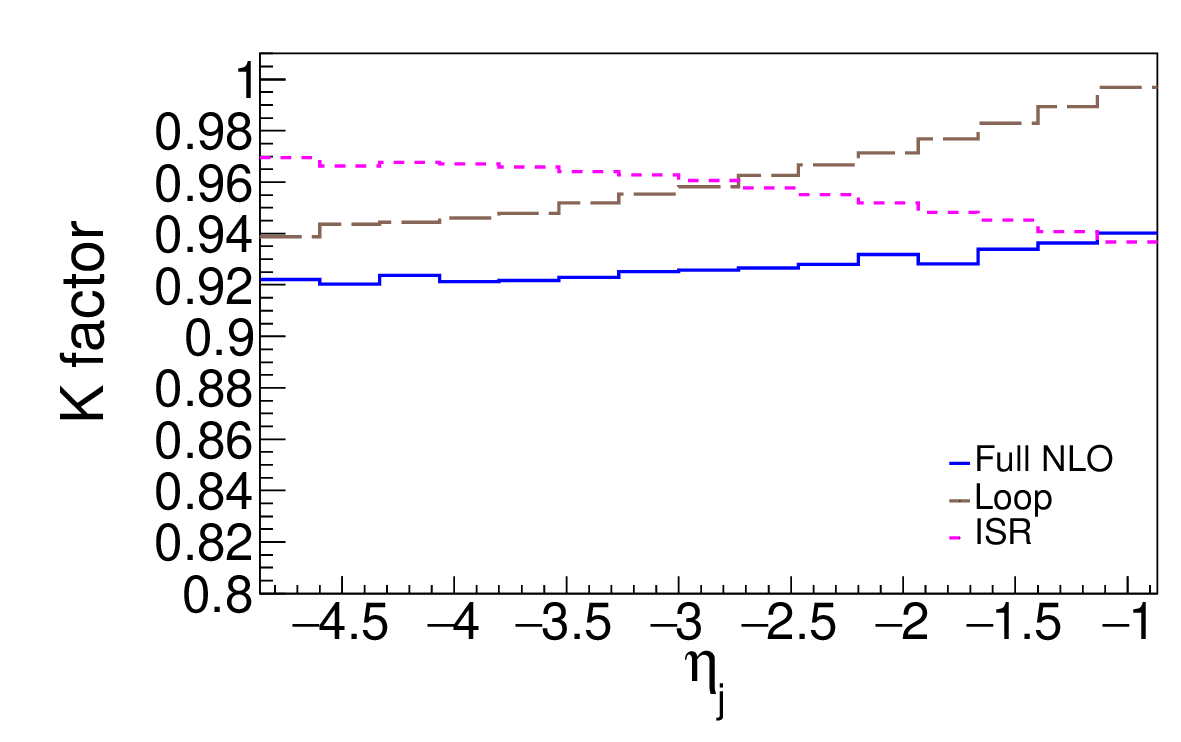}}\qquad \\
	\caption{Distribution in the rapidity $\eta_j$ of the tagging jet (a), and the corresponding $K$ factors (b).}
\label{etaj}
\end{figure}

The tagging jet rapidity in Fig.~\ref{etaj} covers mostly the backward region and peaks around $\eta_{j}= -3$. The relative corrections at 
NLO are still rather stable ($\sim 7\%$) for different $\eta_{j}$, even though the K factors of loop and ISR contributions vary slowly in
opposite trends as $\eta_j$ changes.

\begin{figure}[h!]
\centering
\subfloat[$p_T^{e}$ distribution]{\includegraphics[scale=0.4]{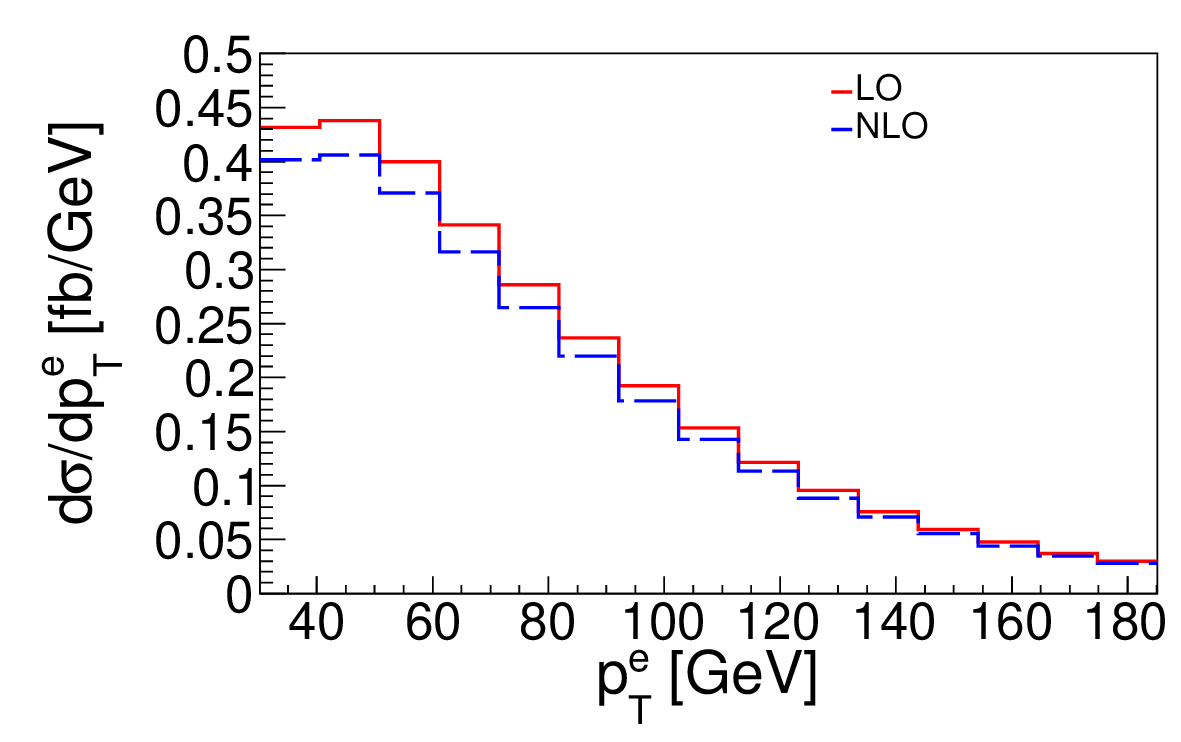}}\qquad
\subfloat[$K$ factor]{\includegraphics[scale=0.4]{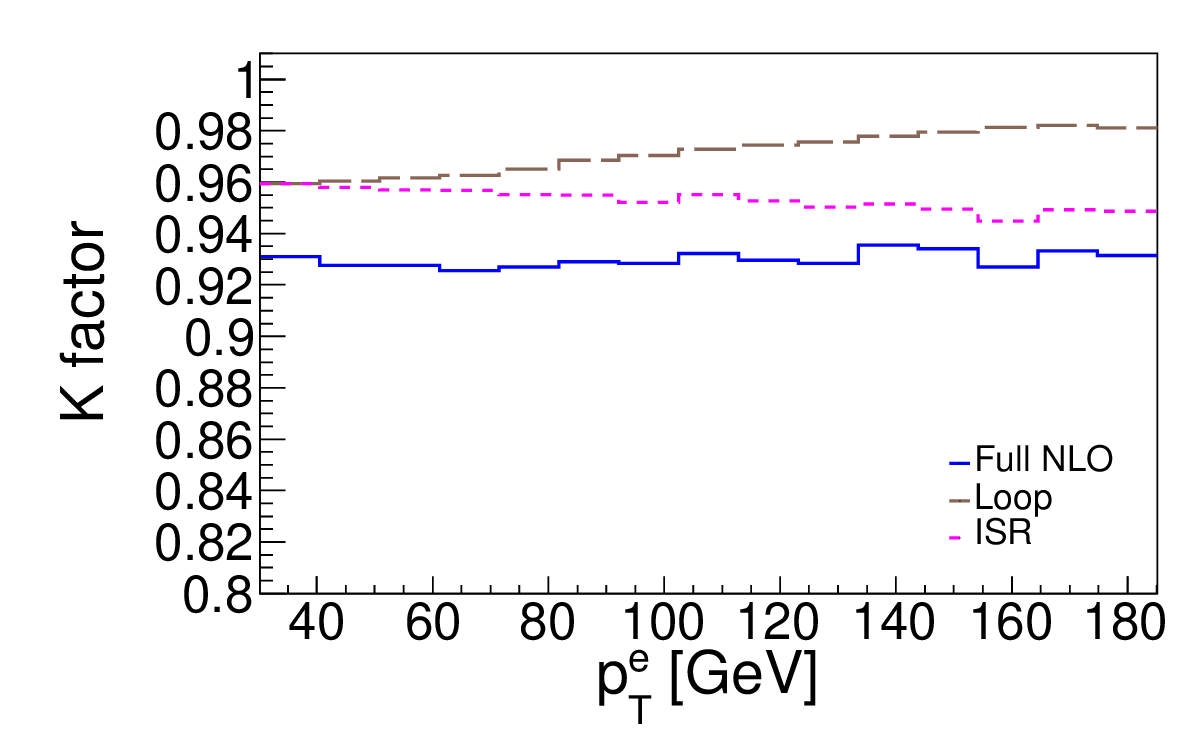}}\qquad \\
	\caption{Distribution in the transverse momentum $p_T^e$ of the outgoing electron (a), and the corresponding $K$ factors (b).}
\label{ptl}
\end{figure}

\begin{figure}[h!]
\centering
\subfloat[$\eta_e$ distribution]{\includegraphics[scale=0.4]{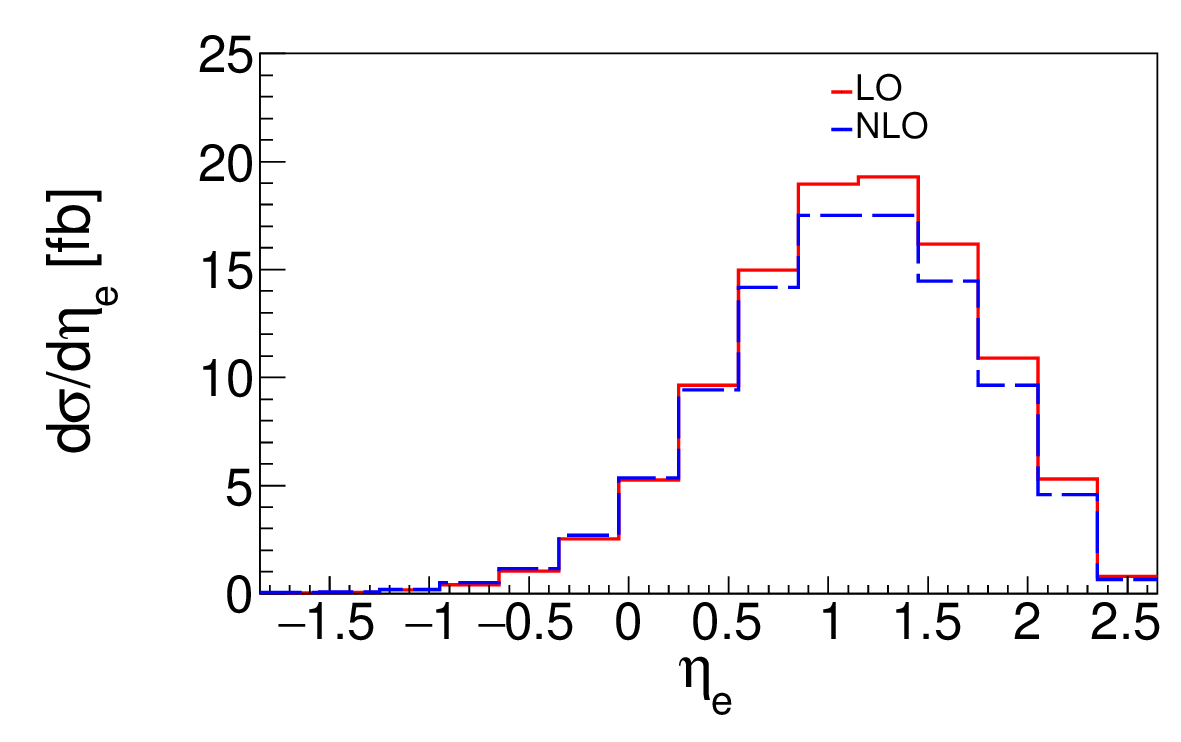}}\qquad
\subfloat[$K$ factor]{\includegraphics[scale=0.4]{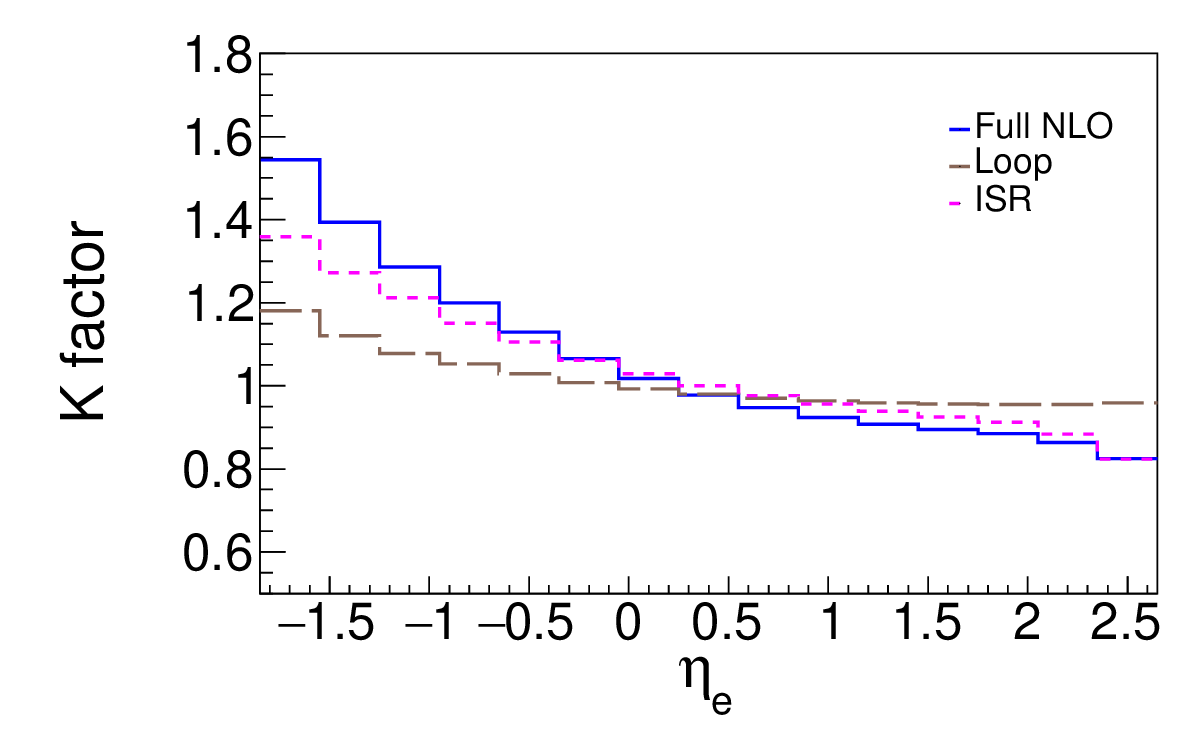}}\qquad \\
	\caption{Distribution in the rapidity $\eta_e$ of the outgoing electron (a), and the corresponding $K$ factors (b).}
\label{etal}
\end{figure}

In searchings of Higgs invisible decays, it is important to make use of the electron observables together with those of the tagging jet.
This proves to be useful in suppressing the background $p + e^- \rightarrow W^-(\rightarrow e^-\bar{\nu}_e) + j + \nu_e$ at the 
LHeC~\cite{Tang:2015uha}.
The distribution of $p_T^e$ in Fig.~\ref{ptl} behaves very similar to that of $p_T^j$, both at LO and NLO, as to be expected from the 
characteristics of WBF events. In contrast, the electron and tagging jet rapidities are very different. The outgoing electron tends to
be in the forward region with a more concentrated distribution of $\eta_e$ (as compared with $\eta_j$) around 1 (see Fig.~\ref{etal}(a)). 
Furthermore, the NLO corrections in Fig.~\ref{etal}(b) become positive and sizeable (as large as 35\%) in going to the negative $\eta_e$, 
despite that the cross sections in this region die out. 
It is interesting to observe that both the loop and ISR corrections at NLO are positive in this region.  While the loop terms are very 
complicated to analyse, it is fairly straightforward to explain the behavior of the ISR corrections. The negative corrections from the virtual ISR
become more important when the CM energy of the beams is near the threshold of 
producing the final state~\cite{Denner:2003ri,Denner:2004jy}. The K factor of the ISR above 1 in the negative $\eta_e$ region, however,
suggests the converse scenario where the CM energy is much larger than the invariant mass of the final state. 
This allows for a large phase space for the positive contribution from the real photon radiations off the inital electron that
may compensate the negative corrections of the virtual ISR. The net NLO corrections at large and positive $\eta_e$ reaches -15\%.
Over the entire spectrum of $\eta_e$, the variation of loop corrections is milder than the ISR.

\begin{figure}[h!]
\centering
\subfloat[$p_T^{H}$ distribution]{\includegraphics[scale=0.4]{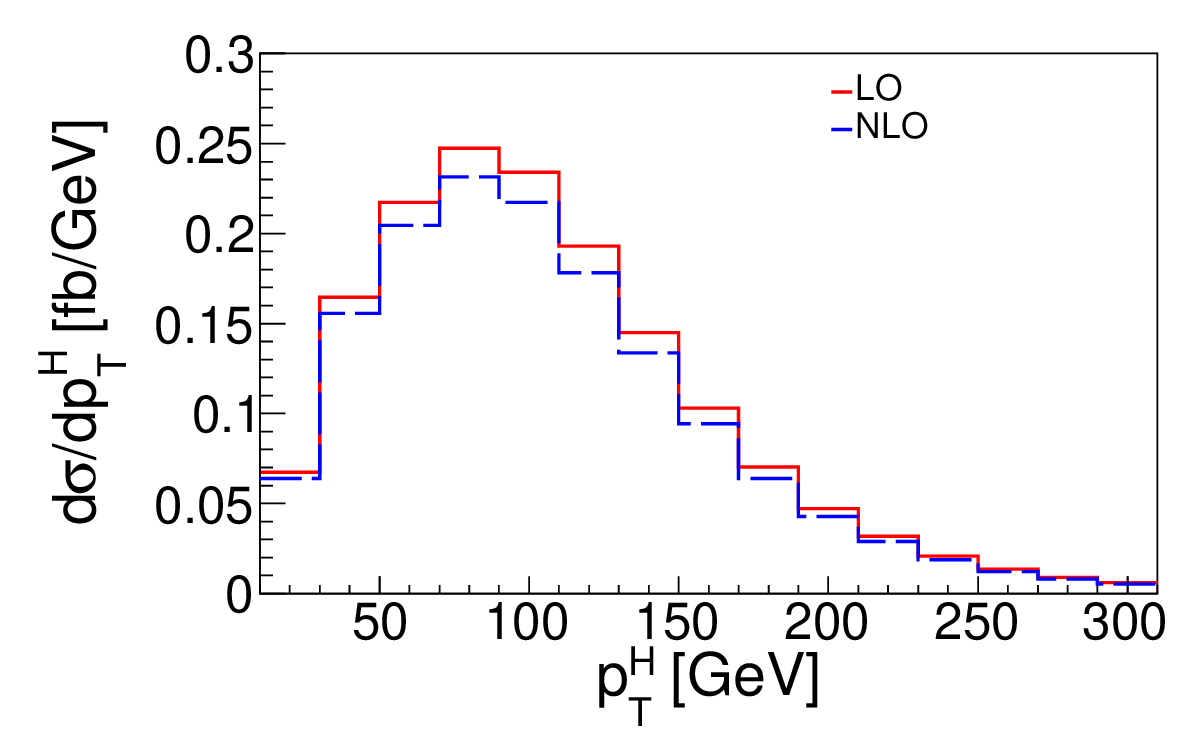}}\qquad
\subfloat[$K$ factor]{\includegraphics[scale=0.4]{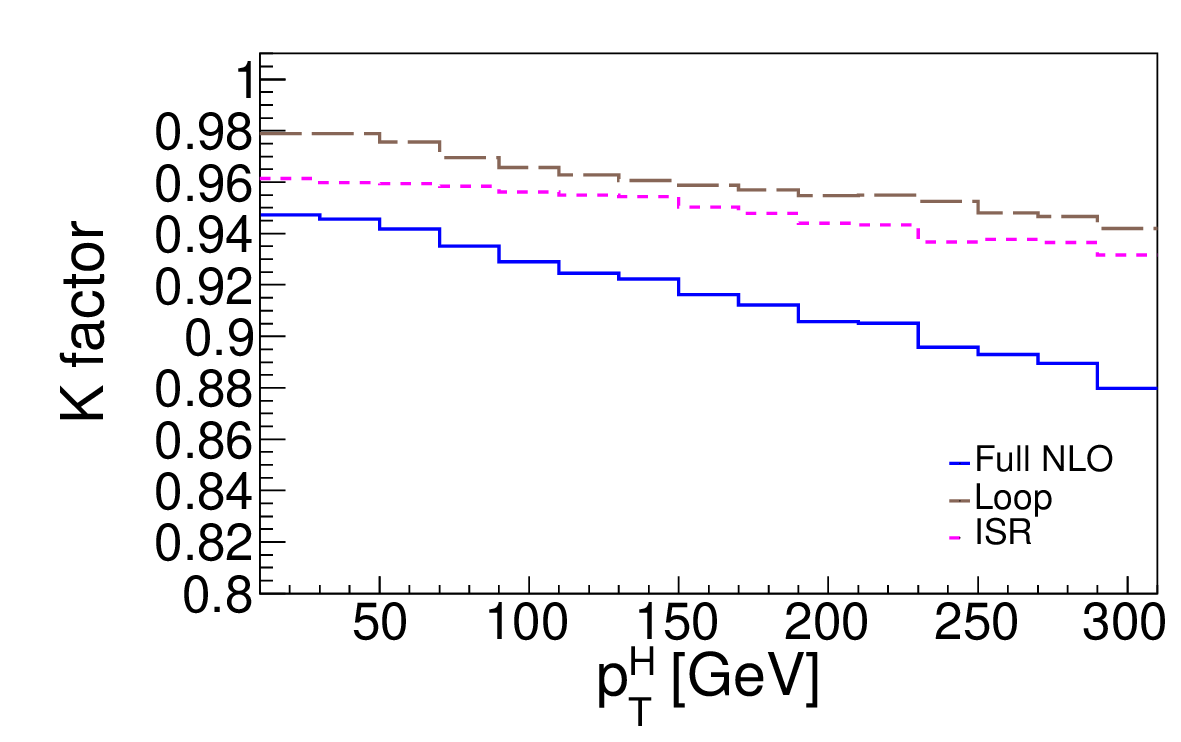}}\qquad \\
	\caption{Distribution in the transverse momentum $p_T^H$ of the Higgs boson (a), and the corresponding $K$ factors (b).}
\label{pth}
\end{figure}

\begin{figure}[h!]
\centering
\subfloat[$\eta_{H}$ distribution]{\includegraphics[scale=0.4]{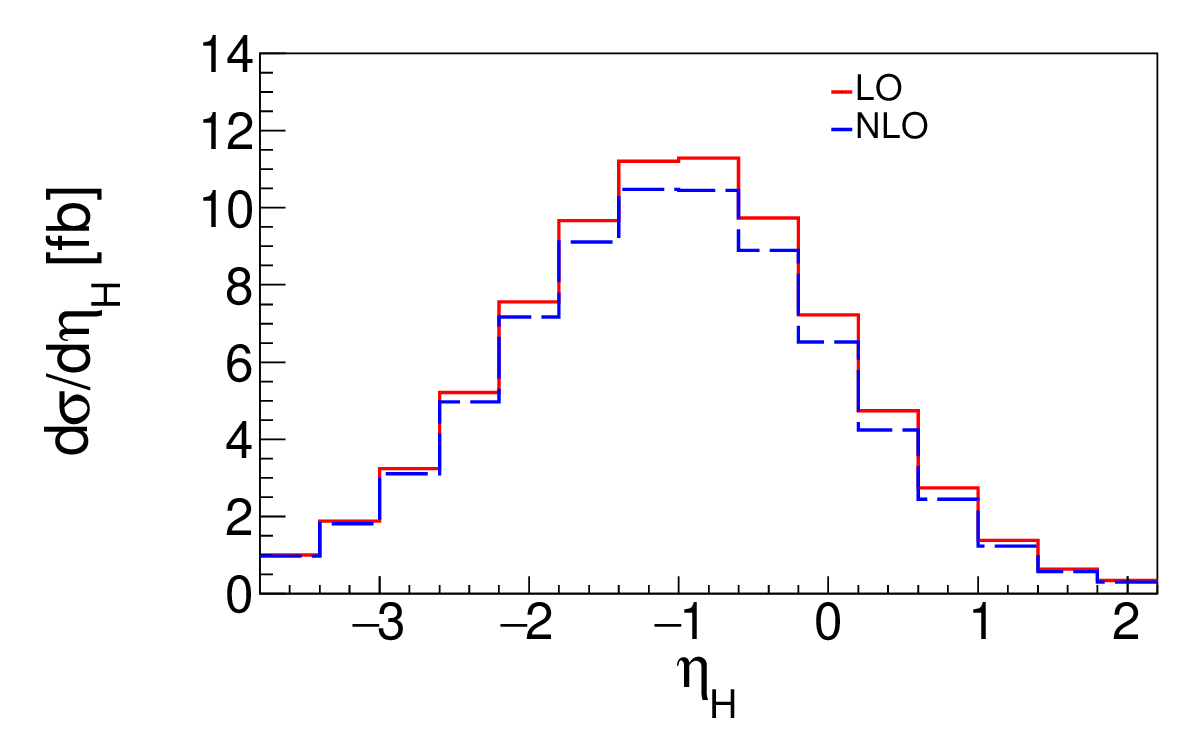}}\qquad
\subfloat[$K$ factor]{\includegraphics[scale=0.4]{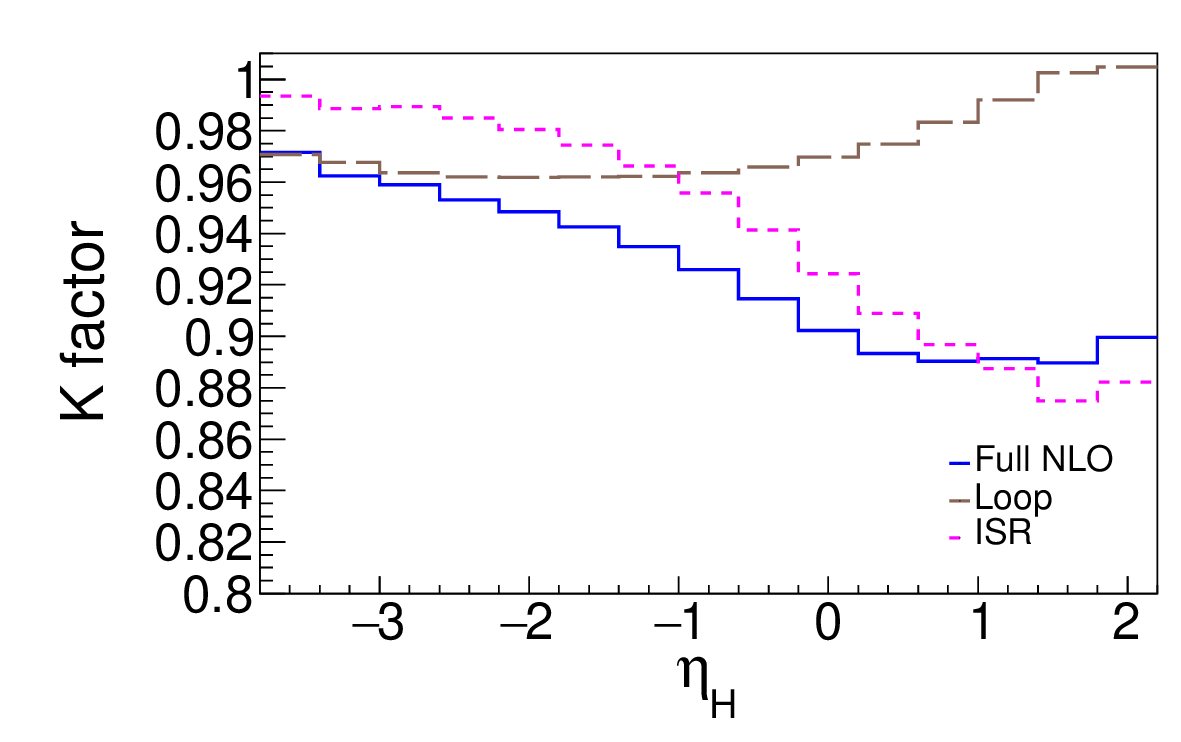}}\qquad \\
	\caption{Distribution in the rapidity $\eta_H$ of the Higgs boson (a), and the corresponding $K$ factors (b).}
\label{etah}
\end{figure}

The typical transverse momentum of the Higgs is much larger than those of the final state fermions. Fig.~\ref{pth} shows that $p_T^H$ peaks at
100 GeV and extends over 300 GeV. The relative corrections at NLO changes between -5\% to -12\% in this range, where the loop and ISR terms 
give comparable contributions. The $\eta_H$ distribution in Fig.~\ref{etah} centers near -1 and appears fairly symmetric on the two sides.
Within the range $-4 < \eta_H <2$, the NLO K factor varies between 0.97 to 0.9, and essentially interpolates between the curves of the loop
terms and of the ISR. The decrease of the ISR K factor at large $p_T^H$ and positive $\eta_H$ is consistent with the fact that these regions
are reached by a large collision energy.

\begin{figure}[h!]
\centering
\subfloat[$\Delta\phi_{e-j}$ distribution]{\includegraphics[scale=0.4]{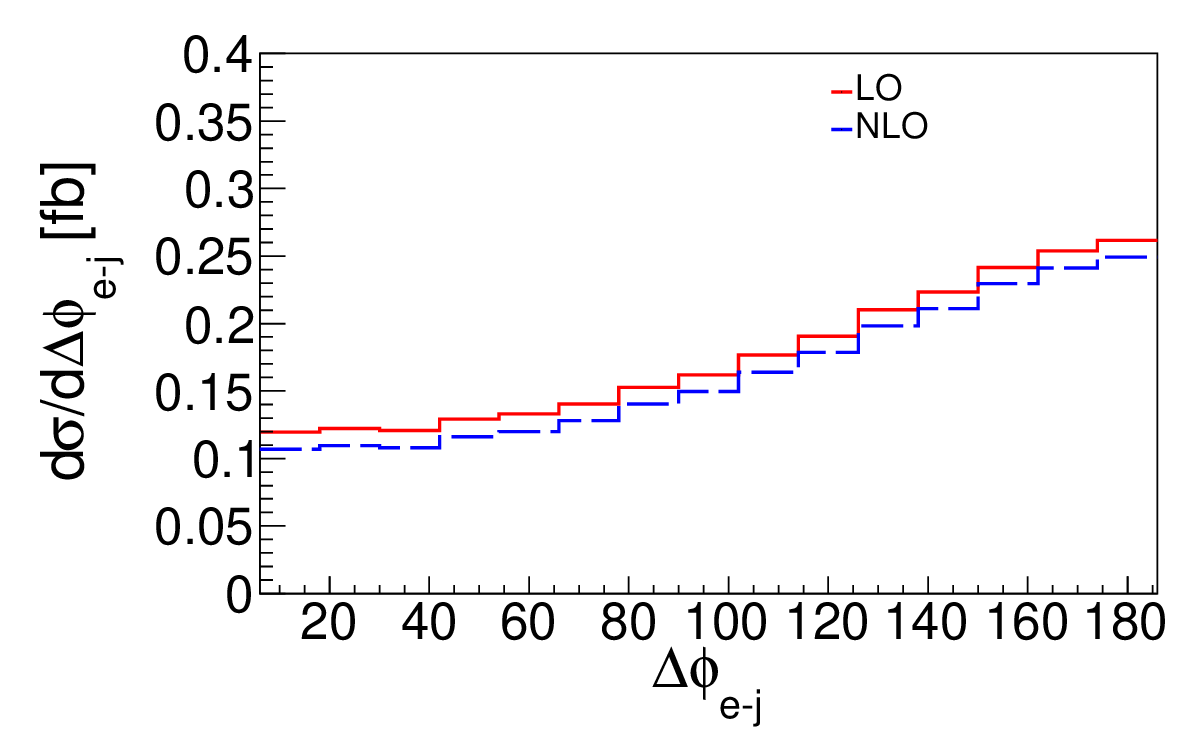}}\qquad
\subfloat[$K$ factor]{\includegraphics[scale=0.4]{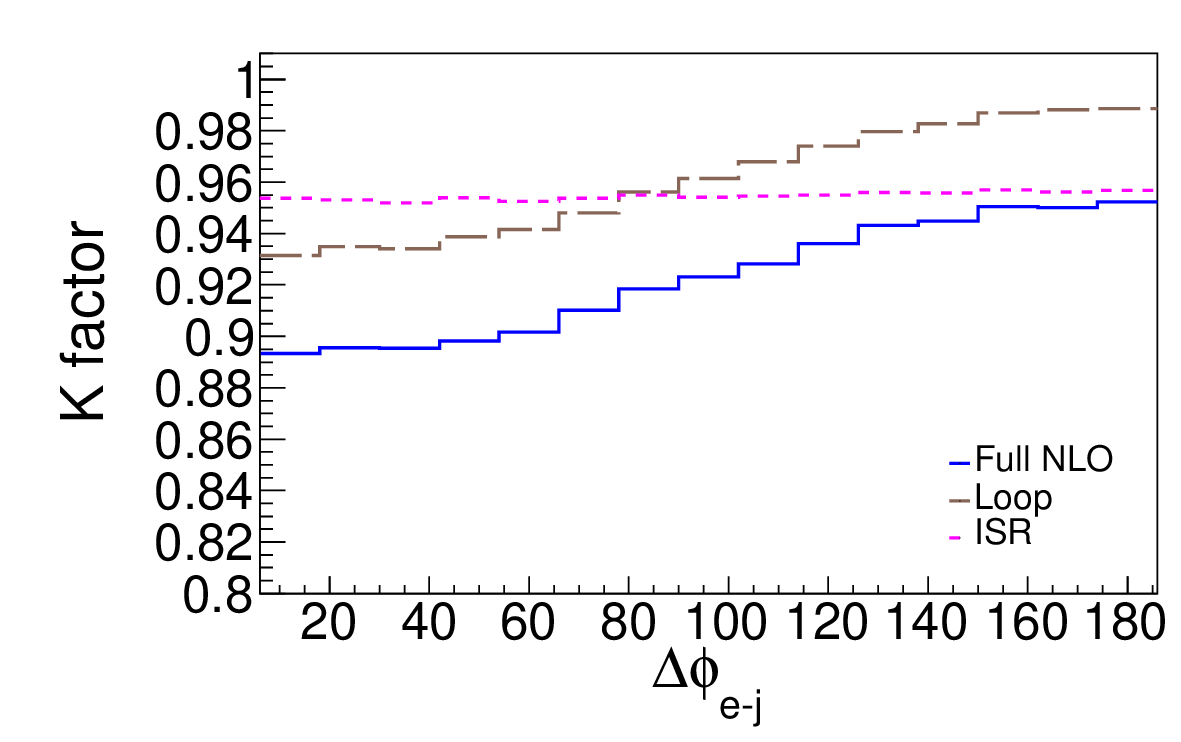}}
	\caption{Distribution in the azimuthal angle difference between the tagging jet and the outgoing electron (a), 
	and the corresponding $K$ factors (b).}
\label{dphi}
\end{figure}

The last observable we shall explore is the azimuthal angle difference between the tagging jet and the final state electron.
This quantity has been shown to be sensitive to Higgs anomalous couplings with gauge bosons, and are used to distinguish these couplings
from the SM ones~\cite{Plehn:2001nj,Biswal:2012mp,Sharma:2022epc}. As with the CC case, the distribution of 
$\Delta\phi_{e-j}$ in Fig.~\ref{dphi} is going through a steady and monotonic increase within the range  $\Delta\phi_{e-j} \in [0,\pi]$.
The K factor increases within the same range from 0.9 to 0.95, with the trend dominated by loop corrections. The ISR terms are not 
sensitive to the change of $\Delta\phi_{e-j}$ and contribute roughly a constant relative correction about -4\%. It is worth mentioning that
the azimuthal angle difference is an effective observable to reduce backgrounds for WBF production of invisible Higgs bosons
at both the LHC~\cite{Eboli:2000ze} and LHeC~\cite{Tang:2015uha}. These searches require the azimuthal angle difference to be 
less than about $70^{\circ}$, which is in Fig.~\ref{dphi} (b) just the region where the NLO corrections are most pronounced.


\section{Conclusion}
\label{conclusion}

In this work we have computed the NLO EW effects for the NC WBF processes at the LHeC. The dipole subtraction formalism is adopted for 
organizing various singularities encountered at 1-loop order. A factorization of the electron distribution is done in order to simplify
the electron mass dependence in the hard scattering cross sections. The numerical result of the calculation is obtained in  
both $G_{\mu}$ and $\alpha(M_Z)$ renormalization schemes for the CM energy of the electron-proton beams at 1.98 TeV. 
While the sizes of the NLO terms computed in the two schemes differ by $\sim 50\%$, 
the results of LO + NLO in two schemes agree up to higher order corrections. The $G_{\mu}$ scheme leads to a
faster convergence of the perturbative calculation, and is used for computing several differential distributions. In the spectra of the considered 
observables where most contributing events reside, the corrections are negative and within the range [-12\%, 0]. The only large positive 
corrections are observed in the region where the rapidity of the final state electron becomes negative, i.e., as it recoils against the
proton beam. The Higgs observables are more sensitive to the EW radiative effects. The roles of the ISR and loop terms are 
quite different in distorting the LO distributions, despite that they both give sizable corrections. 
Overall, the NLO corrections are at the level of 10\% and are significant
in modifying both the normalization and shape of the LO distributions; they should be included in a full phenomenological analysis of the 
NC WBF production of the Higgs at the LHeC.

\section{Acknowledgement}

This work was supported by National Natural Science Foundation of China (12105068). 	
 B.W. and H.X. were also supported by Hangzhou Normal University Start-up Funds.

\appendix

\section{Expressions for the 4-particle final states}
\label{app:4-particle-final-states}

The specific form of the subtraction term for the quark- and anti-quark-initiated processes
\begin{equation}
\begin{aligned}
    e^-(p_a)+q(p_b) \rightarrow e^-(p_1)+\gamma(p_2)+q(p_3)+H(p_4) \nonumber
\end{aligned}
\end{equation}
reads
\begin{equation}
\begin{aligned}
&|\mathcal{M}_{sub}|^2=\mathcal{D}_{e\gamma,q}+\mathcal{D}_{e\gamma}^{e}+\mathcal{D}_{e\gamma}^{q}+\mathcal{D}_{q\gamma,e}+\mathcal{D}_{q\gamma}^{e}+\mathcal{D}_{q\gamma}^{q}+\mathcal{D}_{\gamma,e}^{e}+\mathcal{D}_{\gamma,q}^{e}+\mathcal{D}_{\gamma}^{e,q}+\mathcal{D}_{\gamma,e}^{q}+\mathcal{D}_{\gamma,q}^{q}+\mathcal{D}_{\gamma}^{q,e} \\ 
&\qquad \quad =-\frac{1}{2p_1p_2}\frac{1}{1-y_{13}}8\pi\mu^{2\epsilon}\alpha\left[\frac{2}{1-z_{13}(1-y_{13})}-1-z_{13}-\epsilon(1-z_{13})\right]\boldsymbol{Q}_{13}^2|\mathcal{M}^q_0(\tilde{p}_1,\tilde{p}_3)|^2\\
&\qquad \qquad -\frac{1}{2p_1p_2}\frac{1}{x_{12,a}}8\pi\mu^{2\epsilon}\alpha\left[\frac{2}{2-x_{12,a}-z_{1a}}-1-z_{1a}-\epsilon(1-z_{1a})\right]\boldsymbol{Q}_{1a}^2|\mathcal{M}^q_0(x_{12,a}p_a+P_{1a};x_{12,a}p_a)|^2\\
&\qquad \qquad -\frac{1}{2p_1p_2}\frac{1}{x_{12,b}}8\pi\mu^{2\epsilon}\alpha\left[\frac{2}{2-x_{12,b}-z_{1b}}-1-z_{1b}-\epsilon(1-z_{1b})\right]\boldsymbol{Q}_{1b}^2|\mathcal{M}^q_0(x_{12,b}p_b+P_{1b};x_{12,b}p_b)|^2\\
&\qquad \qquad -\frac{1}{2p_3p_2}\frac{1}{1-y_{31}}8\pi\mu^{2\epsilon}\alpha\left[\frac{2}{1-z_{31}(1-y_{31})}-1-z_{31}-\epsilon(1-z_{31})\right]\boldsymbol{Q}_{13}^2|\mathcal{M}^q_0(\tilde{p}_1,\tilde{p}_3)|^2\\
&\qquad \qquad -\frac{1}{2p_3p_2}\frac{1}{x_{32,a}}8\pi\mu^{2\epsilon}\alpha\left[\frac{2}{2-x_{32,a}-z_{3a}}-1-z_{3a}-\epsilon(1-z_{3a})\right]\boldsymbol{Q}_{3a}^2|\mathcal{M}^q_0(x_{32,a}p_a+P_{3a};x_{32,a}p_a)|^2\\
&\qquad \qquad -\frac{1}{2p_3p_2}\frac{1}{x_{32,b}}8\pi\mu^{2\epsilon}\alpha\left[\frac{2}{2-x_{32,b}-z_{3b}}-1-z_{3b}-\epsilon(1-z_{3b})\right]\boldsymbol{Q}_{3b}^2|\mathcal{M}^q_0(x_{32,b}p_b+P_{3b};x_{32,b}p_b)|^2 \\
&\qquad \qquad -\frac{1}{2p_ap_2}\frac{1}{x_{a2,1}}8\pi\mu^{2\epsilon}\alpha\left[\frac{2}{2-x_{a2,1}-z_{1a}}-1-x_{a2,1}-\epsilon(1-x_{a2,1})\right]\boldsymbol{Q}_{1a}^2|\mathcal{M}^q_0(x_{a2,1}p_a+P_{1a};x_{a2,1}p_a)|^2\\
&\qquad \qquad -\frac{1}{2p_ap_2}\frac{1}{x_{a2,3}}8\pi\mu^{2\epsilon}\alpha\left[\frac{2}{2-x_{a2,3}-z_{3a}}-1-x_{a2,3}-\epsilon(1-x_{a2,3})\right]\boldsymbol{Q}_{3a}^2|\mathcal{M}^q_0(x_{a2,3}p_a+P_{3a};x_{a2,3}p_a)|^2\\
&\qquad \qquad -\frac{1}{2p_ap_2}\frac{1}{x_{a2,b}}8\pi\mu^{2\epsilon}\alpha\left[\frac{2}{1-x_{a2,b}}-1-x_{a2,b}-\epsilon(1-x_{a2,b})\right]\boldsymbol{Q}_{ab}^2|\mathcal{M}^q_0(\tilde{k}_i(x_{a2,b});x_{a2,b}p_a)|^2\\
&\qquad \qquad -\frac{1}{2p_bp_2}\frac{1}{x_{b2,1}}8\pi\mu^{2\epsilon}\alpha\left[\frac{2}{2-x_{b2,1}-z_{1b}}-1-x_{b2,1}-\epsilon(1-x_{b2,1})\right]\boldsymbol{Q}_{1b}^2|\mathcal{M}^q_0(x_{b2,1}p_b+P_{1b};x_{b2,1}p_b)|^2\\
&\qquad \qquad -\frac{1}{2p_bp_2}\frac{1}{x_{b2,3}}8\pi\mu^{2\epsilon}\alpha\left[\frac{2}{2-x_{b2,3}-z_{3b}}-1-x_{b2,3}-\epsilon(1-x_{b2,3})\right]\boldsymbol{Q}_{3b}^2|\mathcal{M}^q_0(x_{b2,3}p_b+P_{3b};x_{b2,3}p_b)|^2\\
&\qquad \qquad -\frac{1}{2p_bp_2}\frac{1}{x_{b2,a}}8\pi\mu^{2\epsilon}\alpha\left[\frac{2}{1-x_{b2,a}}-1-x_{b2,a}-\epsilon(1-x_{b2,a})\right]\boldsymbol{Q}_{ab}^2|\mathcal{M}^q_0(\tilde{k}_i(x_{b2,a});x_{b2,a}p_b)|^2,
\label{eq:realdipq}
\end{aligned}
\end{equation}
where $\mathcal{M}^q_0$ is the amplitude for the born process $e q \rightarrow \nu_e q H$. The momentum fractions are defined as
\begin{equation}
\begin{aligned}
&y_{13}=\frac{p_1p_2}{p_1p_2+p_1p_3+p_2p_3}, \qquad z_{13}=\frac{p_1p_3}{p_1p_3+p_2p_3},\\
&y_{31}=\frac{p_3p_2}{p_1p_2+p_1p_3+p_2p_3}, \qquad z_{31}=\frac{p_1p_3}{p_1p_2+p_1p_3},\\
&x_{12,a}=x_{a2,1}=1-\frac{p_1p_2}{(p_1+p_2)p_a}, \qquad z_{1a}=\frac{p_1p_a}{(p_1+p_2)p_a},\\
&x_{12,b}=x_{b2,1}=1-\frac{p_1p_2}{(p_1+p_2)p_b}, \qquad z_{1b}=\frac{p_1p_b}{(p_1+p_2)p_b},\\
&x_{32,a}=x_{a2,3}=1-\frac{p_3p_2}{(p_3+p_2)p_a}, \qquad z_{3a}=\frac{p_3p_a}{(p_3+p_2)p_a},\\ 
&x_{32,b}=x_{b2,3}=1-\frac{p_3p_2}{(p_3+p_2)p_b}, \qquad z_{3b}=\frac{p_3p_b}{(p_3+p_2)p_b}\\
&x_{a2,b}=x_{b2,a}=1-\frac{p_2(p_a+p_b)}{p_ap_b}.
\label{eq:defx}
\end{aligned}
\end{equation}
As remarked in Sec.~\ref{4-part}, the born amplitudes in Eqs.~\ref{eq:realdipq} depend on the shifted momenta obtained from the 
corresponding radiative processes.
These momenta are shown as the arguments of the born amplitudes, where the momenta from the initial/final state are on the right/left
of the semicolon (Momenta with no modifications are not shown).
They are related to the momenta of the real emission processes by
\begin{equation}
\begin{aligned}
&\tilde{p}_1=p_1+p_2-\frac{y_{13}}{1-y_{13}}p_3 \quad \text{and} \quad \tilde{p}_3=\frac{1}{1-y_{13}}p_3 \quad \text{for} \quad \mathcal{D}_{e\gamma,q'},\\
&\tilde{p}_1=\frac{1}{1-y_{31}}p_1 \quad \text{and} \quad \tilde{p}_3=p_3+p_2-\frac{y_{31}}{1-y_{31}}p_1 \quad \text{for} \quad \mathcal{D}_{q'\gamma,e},\\
&P_{1a}=p_1+p_2-p_a=p_1+p_2-(1-x_{12,a})p_a-x_{12,a}p_a, \\
&P_{1b}=p_1+p_2-p_b=p_1+p_2-(1-x_{12,b})p_b-x_{12,b}p_b, \\
&P_{3a}=p_3+p_2-p_a=p_3+p_2-(1-x_{32,a})p_a-x_{32,a}p_a, \\
&P_{3b}=p_3+p_2-p_b=p_3+p_2-(1-x_{32,b})p_b-x_{32,b}p_b, \\
&\tilde{k}_i^{\mu}=\Lambda_{\nu}^{\mu}k_i^{\nu}=\bigg(g_{\nu}^{\mu}-\frac{(P_{ab}+\tilde{P}_{ab})^{\mu}(P_{ab}+\tilde{P}_{ab})_{\nu}}{P_{ab}^2+P_{ab}\tilde{P}_{ab}}+\frac{\tilde{P}_{ab}^{\mu}P_{ab,\nu}}{P_{ab}^2}\bigg)k_i^{\nu},\\
&P_{ab}=p_a+p_b-p_2 \quad \text{for} \quad \tilde{k}_i(x_{a2,b}) \quad \text{and} \quad \tilde{k}_i(x_{b2,a}),\\
&\tilde{P}_{ab}=x_{a2,b}p_a+p_b \quad \text{for} \quad \tilde{k}_i(x_{a2,b}) \quad \text{and} \quad \tilde{P}_{ab}=p_a+x_{b2,a}p_b \quad \text{for} \quad \tilde{k}_i(x_{b2,a}).
\label{eq:Momtrans}
\end{aligned}
\end{equation}
Note that $\tilde{k}_i$ for all final state particles in $\mathcal{D}_{\gamma}^{e,q}$ and $\mathcal{D}_{\gamma}^{q,e}$ 
must transform according to Eq.~\ref{eq:Momtrans}.
The charge correlator $\boldsymbol{Q}_{ik}^2$ is defined by the charges of the flavor $i$ and $k$ as
\begin{equation}
\begin{aligned}
	&\boldsymbol{Q}_{ik}^2=Q_iQ_k\theta_i\theta_k,
\label{eq:defQ-gen}
\end{aligned}
\end{equation}
where $\theta_{i/k}$ is $1$ ($-1$) if it is in the final (initial) state.
For processes initiated by various quark (antiquark) flavors $q$ that contribute in our calculation, the charge correlators in Eq.~\ref{eq:realdipq} 
are summarized in table~\ref{qiniQ2}.
\begin{table}[h]
\begin{center}
\begin{tabular}{|c|c|c|c|c|c|c|}
\hline
	$q$ & $\boldsymbol{Q}_{13}^2$ & $\boldsymbol{Q}_{1a}^2$ & $\boldsymbol{Q}_{1b}^2$ & $\boldsymbol{Q}_{3a}^2$ & $\boldsymbol{Q}_{3b}^2$ & $\boldsymbol{Q}_{ab}^2$\\
\hline
	$u,\, c$ & $-\frac{2}{3}$ & -1 & $\frac{2}{3}$ & $\frac{2}{3}$ & $-\frac{4}{9}$ & $-\frac{2}{3}$ \\
\hline
	$d,\, s$ & $\frac{1}{3}$ & -1 & $-\frac{1}{3}$ & $-\frac{1}{3}$ & $-\frac{1}{9}$ & $\frac{1}{3}$ \\
\hline
	$\bar{u},\, \bar{c}$ & $\frac{2}{3}$ & -1 & $-\frac{2}{3}$ & $-\frac{2}{3}$ & $-\frac{4}{9}$ & $\frac{2}{3}$ \\
\hline
	$\bar{d},\, \bar{s}$ & $-\frac{1}{3}$ & -1 & $\frac{1}{3}$ & $\frac{1}{3}$ & $-\frac{1}{9}$ & $-\frac{1}{3}$\\
\hline
\end{tabular}
\end{center}
	\caption{Charge correlators of the dipole terms for quark- and antiquark-induced real emission processes.}
\label{qiniQ2}
\end{table}

For the photon-induced processes
\begin{equation}
\begin{aligned}
	e^-(p_a)+\gamma(p_b) \rightarrow e(p_1)+q(p_2)+\bar{q}(p_3)+H(p_4) \nonumber
\end{aligned}
\end{equation}
the subtraction term takes the form
\begin{equation}
\begin{aligned}
	|\mathcal{M}_{sub}|^2&=\mathcal{D}_{q,\bar{q}}^{\gamma}+\mathcal{D}_{q,e}^{\gamma}+\mathcal{D}_{q}^{\gamma,e}+\mathcal{D}_{\bar{q},q}^{\gamma}+\mathcal{D}_{\bar{q},e}^{\gamma}+\mathcal{D}_{\bar{q}}^{\gamma,e}\\
	&=-\frac{1}{2p_bp_2}\frac{1}{x_{b2,3}}8\pi \mu^{2\epsilon} \alpha \Big[1-\epsilon -2x_{b2,3}(1-x_{b2,3})\Big]N_{C,f}\boldsymbol{Q}_{\widetilde{b2}3}^2|\mathcal{M}^{\bar{q}}_0(x_{b2,3}p_b+P_{23};x_{b2,3}p_b)|^2\\
&\quad -\frac{1}{2p_bp_2}\frac{1}{x_{b2,1}}8\pi \mu^{2\epsilon} \alpha \Big[1-\epsilon-2x_{b2,1}(1-x_{b2,1})\Big]N_{C,f} \boldsymbol{Q}_{\widetilde{b2}1}^2| \mathcal{M}^{\bar{q}}_0(x_{b2,1}p_b+P_{21};x_{b2,1}p_b)|^2\\
&\quad -\frac{1}{2p_bp_2}\frac{1}{x_{b2,a}}8\pi \mu^{2\epsilon} \alpha \Big[1-\epsilon -2x_{b2,a}(1-x_{b2,a})\Big]N_{C,f}\boldsymbol{Q}_{\widetilde{b2}a}^2|\mathcal{M}^{\bar{q}}_0(\tilde{k}_i(x_{b2,a});x_{b2,a}p_b)|^2\\
&\quad -\frac{1}{2p_bp_3}\frac{1}{x_{b3,2}}8\pi \mu^{2\epsilon} \alpha \Big[1-\epsilon -2x_{b3,2}(1-x_{b3,2})\Big]N_{C,f}\boldsymbol{Q}_{\widetilde{b3}2}^2|\mathcal{M}^{q}_0(x_{b3,2}p_b+P_{32};x_{b3,2}p_b)|^2\\
&\quad -\frac{1}{2p_bp_3}\frac{1}{x_{b3,1}}8\pi \mu^{2\epsilon} \alpha \Big[1-\epsilon-2x_{b3,1}(1-x_{b3,1})\Big]N_{C,f} \boldsymbol{Q}_{\widetilde{b3}1}^2| \mathcal{M}^{q}_0(x_{b3,1}p_b+P_{31};x_{b3,1}p_b)|^2\\
	&\quad -\frac{1}{2p_bp_3}\frac{1}{x_{b3,a}}8\pi \mu^{2\epsilon} \alpha \Big[1-\epsilon -2x_{b3,a}(1-x_{b3,a})\Big]N_{C,f}\boldsymbol{Q}_{\widetilde{b3}a}^2|\mathcal{M}^{q}_0(\tilde{k}_i(x_{b3,a});x_{b3,a}p_b)|^2,
\label{eq:realdipA}
\end{aligned}
\end{equation}
where $\mathcal{M}^{\bar{q}}_0$ and $\mathcal{M}^{q}_0$ are the amplitudes for the born processes $e \bar{q} \rightarrow e \bar{q} H$ 
and $e q \rightarrow e q H$, respectively. In addition to the definitions made in Eq.~\ref{eq:defx} and ~\ref{eq:Momtrans}, we have
\begin{equation}
\begin{aligned}	
 &x_{b2,3}=x_{b3,2}=1-\frac{p_3p_2}{(p_3+p_2)p_b}, \qquad x_{b2,1}=1-\frac{p_1p_2}{(p_1+p_2)p_b},\qquad x_{b3,1}=1-\frac{p_1p_3}{(p_1+p_3)p_b},\\
 &x_{b2,a}=1-\frac{p_2(p_a+p_b)}{p_ap_b}, \qquad x_{b3,a}=1-\frac{p_3(p_a+p_b)}{p_ap_b}, \qquad N_{C,f}=3, \\
 &P_{23}=P_{32}=p_2+p_3-p_b=p_2+p_3-(1-x_{b2,3})p_b-x_{b2,3}p_b, \\
 &P_{21}=p_1+p_2-p_b=p_1+p_2-(1-x_{b2,1})p_b-x_{b2,1}p_b, \\
 &P_{31}=p_1+p_3-p_b=p_1+p_3-(1-x_{b3,1})p_b-x_{b3,1}p_b, \\
&P_{ab}=p_a+p_b-p_3 \quad \text{and} \quad \tilde{P}_{ab}=p_a+x_{b3,a}p_b \quad \text{for} \quad \tilde{k}_i(x_{b3,a}).
\end{aligned}
\end{equation}
The ``tilde'' symbol in the subscript of each charge correlator denotes the flavor 
from the photon splitting that enters the corresponding born process, 
namely, $\widetilde{b2}$ denotes $\bar{q}$ and $\widetilde{b3}$ denotes $q$.
The charge correlators for choices of $q$ in Eq.~\ref{eq:realdipA} are listed in table~\ref{AiniQ2}
\begin{table}[h]
\begin{center}
\begin{tabular}{|c|c|c|c|c|c|c|}
\hline
	$q$ & $\boldsymbol{Q}_{\widetilde{b2}3}^2$ & $\boldsymbol{Q}_{\widetilde{b2}1}^2$ & $\boldsymbol{Q}_{\widetilde{b2}a}^2$ & $\boldsymbol{Q}_{\widetilde{b3}2}^2$ & $\boldsymbol{Q}_{\widetilde{b3}1}^2$ & $\boldsymbol{Q}_{\widetilde{b3}a}^2$\\ 
\hline
	$u,\, c$ & $-\frac{4}{9}$ & $-\frac{2}{3}$ & $\frac{2}{3}$ & $-\frac{4}{9}$ & $\frac{2}{3}$ & $-\frac{2}{3}$ \\
\hline
	$d,\, s$ & $-\frac{1}{9}$ & $\frac{1}{3}$ & $-\frac{1}{3}$ & $-\frac{1}{9}$ & $-\frac{1}{3}$ & $\frac{1}{3}$ \\
\hline
\end{tabular}
\end{center}
	\caption{Charge correlators of the dipole terms for photon-induced real emission processes.}
\label{AiniQ2}
\end{table}

\section{Expressions for the 3-particle final states}
\label{app:3-particle-final-states}

In this appendix we list the specific form of the $\boldsymbol{I}$,$\boldsymbol{K}$ and $\boldsymbol{P}$ terms 
that are factorized from the born processes
\begin{equation}
\begin{aligned}
    e^-+q \rightarrow e^-(p_1)+q(p_2)+H(p_3), \nonumber
\end{aligned}
\end{equation}
where the momenta of the initial particles are not shown explicitly.
For the corresponding $\boldsymbol{I}$,$\boldsymbol{K}$ and $\boldsymbol{P}$ terms, 
they are given by the arguments of the born factors in the first three lines of Eq.~\ref{eq:virtualsub}.
Recall that there $p_a=P_A$, and $p_b=\eta_bP_B$.

The generic expressions for the quark- and anti-quark-initiated processes are 
\begin{equation}
\begin{aligned}
\boldsymbol{I}^q(\epsilon)&=-\frac{\alpha}{2 \pi} \frac{(4 \pi)^{\epsilon}}{\Gamma(1-\epsilon)}\left[\frac{1}{\epsilon^{2}}+\frac{3}{2 \epsilon}+5-\frac{\pi^{2}}{2}\right] \Bigg\{2\boldsymbol{Q}_{12}^{2}\left(\frac{\mu^{2}}{2 p_{1} p_{2}}\right)^{\epsilon}+2\boldsymbol{Q}_{1a}^{2}\left(\frac{\mu^{2}}{2 p_{a} p_{1}}\right)^{\epsilon}\\
&\qquad \qquad +2\boldsymbol{Q}_{1b}^{2}\left(\frac{\mu^{2}}{2 p_{b}p_{1}}\right)^{\epsilon}+2\boldsymbol{Q}_{2a}^{2}\left(\frac{\mu^{2}}{2p_{a}p_{2}}\right)^{\epsilon}+2\boldsymbol{Q}_{2b}^{2}\left(\frac{\mu^{2}}{2 p_{b} p_{2}}\right)^{\epsilon}+2\boldsymbol{Q}_{ab}^{2}\left(\frac{\mu^{2}}{2 p_{a} p_{b}}\right)^{\epsilon}\Bigg\},
\label{eq:qiniIterm}
\end{aligned}
\end{equation}
\begin{equation}
\begin{aligned}
	\boldsymbol{K}^{c'}_{ff}(x)&=\frac{\alpha}{2\pi}\Bigg\{Q_{f}^2\bigg[-(1+x)\log \frac{1-x}{x}+(1-x)+\Big(\frac{2}{1-x}\log \frac{1-x}{x}\Big)_+ -\delta(1-x)(5-\pi^2)\bigg]\\
	&\quad+\frac{3}{2}\Big(\frac{\boldsymbol{Q}_{1f}^2}{Q_1^2}+\frac{\boldsymbol{Q}_{2f}^2}{Q_2^2}\Big)\bigg[\Big(\frac{1}{1-x}\Big)_++\delta(1-x)\bigg]+\boldsymbol{Q}_{fc'}^2\bigg[(1+x)\log(1-x)-2\bigg(\frac{\log(1-x)}{1-x}\bigg)_+ +\frac{\pi^2}{3}\delta(1-x)\bigg]\Bigg\},
\label{eq:qiniKterm}
\end{aligned}
\end{equation}
\begin{equation}
\begin{aligned}
	\boldsymbol{P}^{c'}_{ff}(x,\mu_F^2)&=\frac{\alpha}{2\pi}\bigg\{-(1+x)+2\Big(\frac{1}{1-x}\Big)_+ +\frac{3}{2}\delta(1-x)\bigg\}\bigg[\boldsymbol{Q}_{f1}^2\log \frac{\mu_F^2}{2xp_fp_1}+\boldsymbol{Q}_{f2}^2\log \frac{\mu_F^2}{2xp_fp_2}+\boldsymbol{Q}_{f{c'}}^2\log \frac{\mu_F^2}{2xp_fp_{c'}}\bigg],
\label{eq:qiniPterm}
\end{aligned}
\end{equation}
where ``$q$'' in $\boldsymbol{I}^q$ can be a quark or anti-quark. In $\boldsymbol{K}^{c'}_{ff}$ and  $\boldsymbol{P}^{c'}_{ff}$, ``$f$''
can be a quark, anti-quark, or electron, while ``$c'$'' is the initial particle species not going through 
splitting (``$c$'' without prime is reserved for the charm quark, which is only a particular case of $c'$. See below). 
The specific choices of charge correlators for Eqs.~\ref{eq:qiniIterm}, ~\ref{eq:qiniKterm}, and ~\ref{eq:qiniPterm} are collected 
in tables~\ref{q-ini-I-term} and ~\ref{q-ini-KP-term}.
\begin{table}[h]
\begin{center}
\begin{tabular}{|c|c|c|c|c|c|c|}
\hline
	$\boldsymbol{I}^q$ & $\boldsymbol{Q}_{12}^2$ & $\boldsymbol{Q}_{1a}^2$ & $\boldsymbol{Q}_{1b}^2$ & $\boldsymbol{Q}_{2a}^2$ & $\boldsymbol{Q}_{2b}^2$ & $\boldsymbol{Q}_{ab}^2$\\
\hline
	$\boldsymbol{I}^{u/c}(\epsilon)$ & $-\frac{2}{3}$ & -1 & $\frac{2}{3}$ & $\frac{2}{3}$ & $-\frac{4}{9}$ & $-\frac{2}{3}$ \\
\hline
	$\boldsymbol{I}^{d/s}(\epsilon)$ & $\frac{1}{3}$ & -1 & $-\frac{1}{3}$ & $-\frac{1}{3}$ & $-\frac{1}{9}$ & $\frac{1}{3}$ \\
\hline
	$\boldsymbol{I}^{\bar{u}/\bar{c}}(\epsilon)$ & $\frac{2}{3}$ & -1 & $-\frac{2}{3}$ & $-\frac{2}{3}$ & $-\frac{4}{9}$ & $\frac{2}{3}$ \\
\hline
	$\boldsymbol{I}^{\bar{d}/\bar{s}}(\epsilon)$ & $-\frac{1}{3}$ & -1 & $\frac{1}{3}$ & $\frac{1}{3}$ & $-\frac{1}{9}$ & $-\frac{1}{3}$\\
\hline
\end{tabular}
\end{center}
	\caption{Charge correlators of $\boldsymbol{I}$ terms for quark- and antiquark-induced processes.}
\label{q-ini-I-term}
\end{table}

\begin{table}[h]
\begin{center}
\begin{tabular}{|c|c|c|c|c|c|c|}
\hline
	$\boldsymbol{R}^{c'}_{ff}$ & $Q_f^2$ & $\boldsymbol{Q}_{1f}^2$ & $\boldsymbol{Q}_{2f}^2$ & $\boldsymbol{Q}_{1}^2$ & $\boldsymbol{Q}_{2}^2$ &$\boldsymbol{Q}_{fc'}^2$\\
\hline
	$\boldsymbol{R}^{u/c}_{ee}$ & 1 & -1 & $\frac{2}{3}$ & 1 & $\frac{4}{9}$ & $-\frac{2}{3}$\\
\hline
	$\boldsymbol{R}^e_{uu/cc}$ & $\frac{4}{9}$ & $\frac{2}{3}$ & $-\frac{4}{9}$ & 1 & $\frac{4}{9}$ & $-\frac{2}{3}$\\
\hline
	$\boldsymbol{R}^{d/s}_{ee}$ & 1 & -1 & $-\frac{1}{3}$ & 1 & $\frac{1}{9}$ & $\frac{1}{3}$\\
\hline
	$\boldsymbol{R}^e_{dd/ss}$ & $\frac{1}{9}$ & $-\frac{1}{3}$ & $-\frac{1}{9}$ & 1 & $\frac{1}{9}$ & $\frac{1}{3}$\\
\hline
	$\boldsymbol{R}^{\bar{u}/\bar{c}}_{ee}$ & 1 & -1 & $-\frac{2}{3}$ & 1 & $\frac{4}{9}$ & $\frac{2}{3}$\\
\hline
	$\boldsymbol{R}^e_{\bar{u}\bar{u}/\bar{c}\bar{c}}$ & $\frac{4}{9}$ & $-\frac{2}{3}$ & $-\frac{4}{9}$ & 1 & $\frac{4}{9}$ & $\frac{2}{3}$\\
\hline
	$\boldsymbol{R}^{\bar{d}/\bar{s}}_{ee}$ & 1 & -1 & $\frac{1}{3}$ & 1 & $\frac{1}{9}$ & $-\frac{1}{3}$\\
\hline
	$\boldsymbol{R}^e_{\bar{d}\bar{d}/\bar{s}\bar{s}}$ & $\frac{1}{9}$ & $\frac{1}{3}$ & $-\frac{1}{9}$ & 1 & $\frac{1}{9}$ & $-\frac{1}{3}$\\
\hline
\end{tabular}
\end{center}
	\caption{Charge correlators of $\boldsymbol{R}$ terms for quark- and antiquark-induced processes, where $\boldsymbol{R}$ stands for $\boldsymbol{K}$ or $\boldsymbol{P}$. Note that
	$\boldsymbol{Q}_{f1}^2$ and $\boldsymbol{Q}_{f2}^2$ are not listed since $\boldsymbol{Q}_{f1}^2=\boldsymbol{Q}_{1f}^2$, and
	$\boldsymbol{Q}_{f2}^2=\boldsymbol{Q}_{2f}^2$.}
\label{q-ini-KP-term}
\end{table}

For the photon-initiated processes, $\boldsymbol{I}$ terms do not contribute. 
Also the electron splitting terms are of higher order in $\alpha$ and do not contribute in this case either. $\boldsymbol{K}$ and $\boldsymbol{P}$ terms take the form
\begin{equation}
\begin{aligned}
	\boldsymbol{K}^a_{\gamma f}(x)&=\frac{\alpha}{2\pi}\Bigg\{N_{C,f}Q_f^2\frac{1+(1-x)^2}{x}\log\frac{1-x}{x}+N_{C,f}Q_f^2x(1-x)-\boldsymbol{Q}_{fa}^2N_{C,f}\frac{1+(1-x)^2}{x}\log(1-x)\Bigg\},
\label{eq:AiniKterm}
\end{aligned}
\end{equation}
\begin{equation}
\begin{aligned}
	\boldsymbol{P}^a_{\gamma f}(x,\mu_F^2)=\frac{\alpha}{2\pi}\bigg\{N_{C,f}\frac{1+(1-x)^2}{x}\bigg\}\bigg[\boldsymbol{Q}_{f1}^2\log \frac{\mu_F^2}{2xp_bp_1}+\boldsymbol{Q}_{f2}^2\log \frac{\mu_F^2}{2xp_bp_2}+\boldsymbol{Q}_{fa}^2\log \frac{\mu_F^2}{2xp_ap_b}\bigg].
\label{eq:AiniPterm}
\end{aligned}
\end{equation}
The charge correlators for Eqs.~\ref{eq:AiniKterm} and ~\ref{eq:AiniPterm} are collected 
in table~\ref{A-ini-KP-term}.
\begin{table}[h]
\begin{center}
\begin{tabular}{|c|c|c|c|c|}
\hline
	$\boldsymbol{R}^a_{\gamma f}$ & $Q_f^2$ & $\boldsymbol{Q}_{f1}^2$ & $\boldsymbol{Q}_{f2}^2$ & $\boldsymbol{Q}_{fa}^2$ \\
\hline
	$\boldsymbol{R}^e_{\gamma u/c}$ & $\frac{4}{9}$ & $\frac{2}{3}$ & $-\frac{4}{9}$ & $-\frac{2}{3}$ \\
\hline
	$\boldsymbol{R}^e_{\gamma d/s}$ & $\frac{1}{9}$ & $-\frac{1}{3}$ & $-\frac{1}{9}$ & $\frac{1}{3}$ \\
\hline
	$\boldsymbol{R}^e_{\gamma \bar{u}/\bar{c}}$ & $\frac{4}{9}$ & $-\frac{2}{3}$ & $-\frac{4}{9}$ & $\frac{2}{3}$ \\
\hline
	$\boldsymbol{R}^e_{\gamma \bar{d}/\bar{s}}$ & $\frac{1}{9}$ & $\frac{1}{3}$ & $-\frac{1}{9}$ & $-\frac{1}{3}$ \\
\hline
\end{tabular}
\end{center}
	\caption{Charge correlators of $\boldsymbol{R}$ terms for photon-induced processes, where $\boldsymbol{R}$ stands for $\boldsymbol{K}$ or $\boldsymbol{P}$.}
\label{A-ini-KP-term}
\end{table}

\bibliographystyle{utphysmcite}
\bibliography{WBF}

\end{document}